\documentclass[useAMS,usenatbib,usegraphicx]{mn2e}
\def\figscl{0.9}

\usepackage{color}

\usepackage[colorlinks,citecolor=black,linkcolor=black,urlcolor=black]{hyperref}
\newcommand*{\figref}[1]{Figure~\ref{#1}}

\newcommand*{\tabref}[1]{Table~\ref{#1}}

\newcommand*{\eqnref}[1]{Equation~\ref{#1}}
\newcommand*{\eqnrefs}[1]{Equations~\ref{#1}}
\newcommand*{\secref}[1]{Section~\ref{#1}}
\newcommand*{\secrefs}[1]{Sections~\ref{#1}}

\newcommand*{\unit}[1]{\ensuremath{\mathrm{\, #1}}}
\newcommand*{\Msun}{\ensuremath{\, M_{\odot}}}
\newcommand*{\erg}{\unit{erg}}
\newcommand*{\keV}{\unit{keV}}
\newcommand*{\second}{\unit{s}}

\newcommand*{\km}{\unit{km}}
\newcommand*{\Mpc}{\unit{Mpc}}

\newcommand*{\E}[1]{\ensuremath{\times 10^{#1}}}
\newcommand*{\logTen}{\ensuremath{\log_{10}}}
\newcommand*{\expectation}[1]{\ensuremath{\left\langle #1 \right\rangle}}
\newcommand*{\like}{\ensuremath{\mathcal{L}}}

\newcommand*{\mysub}[2]{\ensuremath{#1_{\mathrm{#2}}}}
\newcommand*{\mysup}[2]{\ensuremath{#1^{\mathrm{#2}}}}
\newcommand*{\Omegam}{\mysub{\Omega}{m}}
\newcommand*{\Omegab}{\mysub{\Omega}{b}}
\newcommand*{\Omegac}{\mysub{\Omega}{c}}
\newcommand*{\Omegal}{\ensuremath{\Omega_{\Lambda}}}

\newcommand*{\LCDM}{\ensuremath{\Lambda}CDM}

\newcommand*{\fgas}{\mysub{f}{gas}}
\newcommand*{\Mgas}{\mysub{M}{gas}}
\newcommand*{\Mtot}{\mysub{M}{tot}}
\newcommand*{\rhoc}{\mysub{\rho}{cr}}
\newcommand*{\rhom}{\mysub{\bar{\rho}}{m}}
\newcommand*{\wet}{\mysub{w}{et}}

\defcitealias{Allen08}{A08}
\defcitealias{Mantz08}{M08}
\defcitealias{Mantz09a}{Paper~II}
\newcommand*{\scalingpaper}{\citetalias{Mantz09a}}

\newcommand*{\Chandra}{{\it{Chandra}}}
\newcommand*{\ROSAT}{{\it{ROSAT}}}
\newcommand*{\WMAP}{{\it{WMAP}}}
\newcommand*{\Nobs}{\mysub{N}{det}}
\newcommand*{\nobs}{\mysub{\tilde{n}}{det}}
\newcommand*{\Nmis}{\mysub{N}{mis}}

\title[Cluster growth: methods and cosmology]{The Observed Growth of Massive Galaxy Clusters I: Statistical Methods and Cosmological Constraints}
\author[A. Mantz et al.]{
  A.~Mantz,$^{1,2}$\thanks{E-mail: amantz@slac.stanford.edu} S.~W.~Allen,$^{1,2}$ D.~Rapetti$^{1,2}$ and H.~Ebeling$^3$\\
  $^1$Kavli Institute for Particle Astrophysics and Cosmology, Stanford University, 452 Lomita Mall, Stanford, CA 94305-4085, USA\\
  $^2$SLAC National Accelerator Laboratory, 2575 Sand Hill Road, Menlo Park, CA 94025, USA\\
  $^3$Institute for Astronomy, 2680 Woodlawn Drive, Honolulu, HI 96822, USA
}
\date{Accepted 2010 April 28. Received 2010 March 13; in original form 2009 August 30}

\begin{document}
\pagerange{\pageref{firstpage}--\pageref{lastpage}} \pubyear{2010}
\maketitle
\label{firstpage}

\begin{abstract}
  This is the first of a series of papers in which we derive simultaneous constraints on cosmological parameters and X-ray scaling relations using observations of the growth of massive, X-ray flux-selected galaxy clusters. Our data set consists of 238 cluster detections from the \ROSAT{} All-Sky Survey, and incorporates follow-up observations of 94 of those clusters using the \Chandra{} X-ray Observatory or \ROSAT{}. Here we describe and implement a new statistical framework required to self-consistently produce simultaneous constraints on cosmology and scaling relations from such data, and present results on models of dark energy. In spatially flat models with a constant dark energy equation of state, $w$, the cluster data yield $\Omegam=0.23 \pm 0.04$, $\sigma_8=0.82 \pm 0.05$, and $w=-1.01 \pm 0.20$, incorporating standard priors on the Hubble parameter and mean baryon density of the Universe, and marginalizing over conservative allowances for systematic uncertainties. These constraints agree well and are competitive with independent data in the form of cosmic microwave background anisotropies, type Ia supernovae, cluster gas mass fractions, baryon acoustic oscillations, galaxy redshift surveys, and cosmic shear. The combination of our data with current microwave background, supernova, gas mass fraction, and baryon acoustic oscillation data yields $\Omegam=0.27 \pm 0.02$, $\sigma_8=0.79 \pm 0.03$, and $w=-0.96 \pm 0.06$ for flat, constant $w$ models. The combined data also allow us to investigate evolving $w$ models. Marginalizing over transition redshifts in the range 0.05--1, we constrain the equation of state at late and early times to be respectively $w_0=-0.88 \pm 0.21$ and $\wet=-1.05^{+0.20}_{-0.36}$, again including conservative systematic allowances. The combined data provide constraints equivalent to a Dark Energy Task Force figure of merit of 15.5. Our results highlight the power of X-ray studies, which enable the straightforward production of large, complete, and pure cluster samples and admit tight scaling relations, to constrain cosmology. However, the new statistical framework we apply to this task is equally applicable to cluster studies at other wavelengths.
\end{abstract}


\section{Introduction} \label{sec:introduction}

Clusters of galaxies have a long history as cosmological laboratories, beginning with the discovery of dark matter \citep{Zwicky37} through studies of cluster galaxy orbits, and providing early evidence for a low matter density universe \citep{White93a} through studies of their baryonic and dark matter content. These breakthroughs rested on observations of the properties of individual systems, but the population of clusters as a whole also contains a great deal of cosmological information. Clusters represent the most massive gravitationally bound systems in the Universe, and as such their abundance probes the amount of structure in the Universe and its growth over cosmic time.

The local cluster population has been used to jointly constrain the average density of matter, $\Omegam$, and the amplitude of density perturbations, $\sigma_8$ \citep[recently by][but see also \citealt{Reiprich02,Allen03,Schuecker03} and additional references in \citealt{Mantz08}, hereafter \citetalias{Mantz08}]{Henry09,Rozo10}. Most recently, the construction of X-ray flux-limited cluster samples out to redshift $z=0.5$ and beyond has enabled studies of the growth of structure \citep[\citetalias{Mantz08};][]{Vikhlinin09a}. Such studies provide an important new window on cosmology, as various dark energy and modified gravity models designed to explain the acceleration of the cosmic expansion are potentially distinguishable by their effects on structure formation (e.g. \citealt{Rapetti09,Rapetti09a}; \citealt*{Schmidt09}).

The key ingredients for investigations of the cluster population are (1) a cluster survey with a well understood selection function,\footnote{Throughout this paper, we use the term ``selection function'' to refer to the probability that a cluster be detected by a survey and included in the resulting data set, as a function of that cluster's physical properties (redshift, flux etc.). See \secref{sec:likelihood}.} and (2) scaling relations that link cluster mass to observable quantities. Currently, the most successful approach to finding massive clusters over a range of redshifts is through the X-ray emission of the hot intracluster gas, although Sunyaev-Zel'dovich and optical surveys are also making significant progress. Because Malmquist and Eddington biases are ubiquitous in current X-ray flux-limited samples, great care is warranted in their analysis. In particular, precise, robust cosmological constraints can only be obtained by simultaneously fitting the X-ray luminosity--mass relation, as we do here. Conversely, rigorous analysis of the scaling relations must take into account the cluster mass function and the selection function of the data set, as discussed by \citet[][hereafter \scalingpaper{}; see also \citealt{Stanek06,Pacaud07}]{Mantz09a}.

While a flux-limited sample, combined with more detailed follow-up observations of a subset of clusters in that sample, may contain the necessary information to provide simultaneous constraints on cosmology and scaling relations, a fully self-consistent statistical framework for this analysis has been lacking to this point. By ``self-consistent'', we mean that a single likelihood function applying to the full data set (survey + follow-up observations) and encompassing the entire theoretical model (cosmology + scaling relations) should be derived from first principles, ensuring that the covariance among all the model parameters is fully captured and that the effects of the mass function and selection biases are properly accounted for throughout.

This is the first of a series of papers in which we address these issues. \scalingpaper{} contains details of the follow-up X-ray observations and their reduction, presents the constraints on scaling relations from our simultaneous analysis, and discusses their astrophysical implications. In this paper, we present the statistical methods applied to the problems described above and the resulting cosmological constraints. The dark energy models we address here include the simple cosmological constant model; models with a constant dark energy equation of state, $w$; and simple evolving $w$ models. As we will show, our analysis produces some of the tightest constraints on dark energy parameters of any experiment to date. In Papers~III \citep{Rapetti09a} and IV \citep*{Mantz09b}, we respectively apply our analysis to investigations of modified gravity and neutrino properties.

In \secref{sec:data}, we briefly review the cluster data set, which is more fully described in \scalingpaper{}. \secref{sec:modeling} contains a full description of the theoretical model fit to the data, including the cosmology, cluster mass function, and scaling relations. The new, self-consistent analysis method is presented in \secref{sec:likelihood}. \secrefs{sec:results} and \ref{sec:systematics} respectively contain the cosmological results of our analysis and a discussion of relevant systematics.

In this paper, we adopt the conventional definition of cluster radius in terms of the critical density of the Universe; thus, $r_\Delta$ is the radius within which the mean density of the cluster is $\Delta$ times the critical density at the cluster's redshift, $\rhoc(z)$. An alternative convention, used particularly in the literature relevant to the mass function, is to define the overdensity with respect to the mean matter density at redshift $z$. We will consistently use the former convention, and so, for example, an overdensity of 300 with respect to the matter density will be written as $\Delta=300\Omegam(z)$, where $\Omegam(z)$ is the ratio of the mean matter density to the critical density.

\section{Data} \label{sec:data}

The galaxy cluster data used in this work, as well as their selection and reduction, are discussed in detail in \scalingpaper{}. Using three wide-area cluster samples drawn from the \ROSAT{} All-Sky Survey \citep[RASS;][]{Trumper93} -- the \ROSAT{} Brightest Cluster Sample \citep[BCS;][]{Ebeling98}, the \ROSAT{}-ESO Flux-Limited X-ray sample \citep[REFLEX;][]{Bohringer04}, and the bright sub-sample of the Massive Cluster Survey (Bright MACS; \citealt*{Ebeling01}; \citealt{Ebeling10}) -- we select a statistically complete sample of 238 X-ray luminous clusters covering the redshift range $z<0.5$. Of these 238 clusters, 94 have follow-up \Chandra{} or \ROSAT{} observations that we incorporate into the analysis. To distinguish it from cluster data used in other cosmological work (e.g. optically selected clusters), we refer to this data set as the cluster X-ray Luminosity Function (XLF), although in fact the data set contains a great deal more information than the luminosity function alone. From the follow-up observations, we measure X-ray luminosity, average temperature, and gas mass within $r_{500}$. The gas mass is used as a proxy for total mass, using the finding of \citet[][hereafter \citetalias{Allen08}]{Allen08} that the gas mass fraction, $\fgas=\Mgas/\Mtot$, is a constant for the hot, massive clusters being studied.\footnote{We prefer to use $\Mgas$ as a proxy for total mass rather than the thermal energy, $Y_X=\Mgas kT$, because \Mgas{} can be measured more precisely than temperature for a given exposure time, and because \Mgas{} measurements at $r_{500}$ are minimally affected by background and emission weighting uncertainties. In addition, \Mgas{} displays exceptionally small scatter with mass (see \citetalias{Allen08} and \scalingpaper{}) in the mass range considered here, $M>3\E{14}\Msun$.} The systematic uncertainty associated with the \fgas{} measurement is accounted for by simultaneously constraining the full \fgas{} model of \citetalias{Allen08}, which includes generous systematic allowances for instrument calibration, non-thermal pressure support, the depletion of baryons in clusters relative to the cosmic mean, and evolution in the baryon depletion and stellar content of clusters. We use only the six lowest redshift clusters from \citetalias{Allen08} ($z<0.15$), which are sufficient to constrain the gas mass fraction at low redshift without directly producing a constraint on dark energy by themselves. In our analysis of the XLF data, we also use Gaussian priors to constrain the Hubble parameter and the mean baryon density, based on the results of the Hubble Key project \citep[$h=H_0/100\km\second^{-1}\Mpc^{-1}=0.72 \pm 0.08$,][]{Freedman01} and big bang nucleosynthesis studies \citep[$\Omegab h^2=0.0214 \pm 0.002$,][]{Kirkman03}.

In addition, we compare and combine results from our own analysis with those of independent cosmological data, including the full cluster \fgas{} data set \citepalias[][42 clusters at $z<1.1$]{Allen08}, as well as cosmic microwave background (CMB), type Ia supernova (SNIa) and baryon acoustic oscillation (BAO) data. Our analysis of the CMB anisotropies uses 5-year {\it Wilkinson Microwave Anisotropy Probe} (\WMAP{}) data \citep{Hinshaw09,Hill09,Nolta09} with the March 2008 version of the \WMAP{} likelihood code\footnote{\url{http://lambda.gsfc.nasa.gov}} \citep{Dunkley09}. The SNIa results are derived from the Union compilation \citep{Kowalski08}, which includes data from a variety of sources \citep[307 SNIa in total;][]{Hamuy96,Garnavich98,Riess98,Riess99,Riess04,Riess07,Schmidt98,Perlmutter99,Krisciunas01,Krisciunas04,Krisciunas04a,Knop03,Tonry03,Barris04,Astier06,Jha06,Miknaitis07,Wood-Vasey07}, including the treatment of systematic errors employed by Kowalski et~al. Our analysis of BAO data uses the constraints on the ratio of the sound horizon to the distance scale at $z=0.25$ and $z=0.35$ derived by \citet{Percival07} from the galaxy correlation function in 2dF \citep{Colless01,Colless03} and Sloan Digital Sky Survey \citep{Adelman-McCarthy07} data.

When fitting CMB data, we allow the scalar spectral index, $\mysub{n}{s}$, and optical depth to reionization, $\tau$, to vary as free parameters, and marginalize over a plausible range in the amplitude of the Sunyaev-Zel'dovich signal due to galaxy clusters ($0<\mysub{A}{SZ}<2$; introduced by \citealt{Spergel07}). The combination of CMB and \fgas{} data places tight constraints on both $h$ and $\Omegab h^2$ in addition to other parameters of interest (see \citetalias{Allen08}), so the Hubble Key project and big bang nucleosynthesis priors are unnecessary in analyses of the combined data sets. This applies as well to the combination of CMB and XLF data, since we always use the $z<0.15$ subset of the \fgas{} data to calibrate the cluster mass scale for the XLF analysis.

\section{Model} \label{sec:modeling}

In this section, we detail the various pieces of the model being fit to the XLF data. \secref{sec:cosmodel} reviews the important cosmological parameters and the dark energy models to be addressed, while \secref{sec:massfunction} focuses on the predicted mass function and its cosmological dependence. In \secref{sec:scalingmodel}, we outline the model for the cluster scaling relations. \secrefs{sec:samplingmodel} and \ref{sec:surveymodel} describe the sampling model, i.e. the model for how a set of parameters predicts the measured data, including measurement errors and their correlation, and implicit dependences on reference cosmological values. The model parameters are summarized in \tabref{tab:paramlist}.

\begin{table*}
  \centering
  \caption{Parameters and priors used in the analysis. The nuisance parameters associated with the \fgas{} model \citepalias{Allen08} are not shown, although they are also marginalized over. When no entry appears in the prior column, the prior was uniform and significantly wider than the marginal posterior for that parameter. $\mathcal{N}(\mu,\sigma)$ represents the normal distribution with mean $\mu$ and variance $\sigma^2$, and $\mathcal{U}(x_1,x_2)$ the uniform distribution with endpoints $x_1$ and $x_2$. For brevity, $\mathcal{N}_4(\mu,\ldots)$ represents the multivariate normal prior used for the \citet{Tinker08} mass function parameters, where $\mu$ is the marginal mean for each parameter and the covariance matrix is not explicitly shown. Notes: $^a$~when using CMB data, the angular size of the sound horizon at last scattering replaces $h$ as a free parameter; $^b$~indicates priors that are not used when CMB data is included in the analysis; $^c$~indicates parameters that are free only when CMB data is included; $^d$~indicates that there is an independent parameter of this type for each cluster sample.}
  \begin{tabular}{lcll}
    \hline
    Type & Symbol & Meaning & Prior \\
    \hline
    Cosmology (\S\ref{sec:cosmodel})
                     & $h$               & Hubble parameter$~^{a,b}$ & $\mathcal{N}(0.72, 0.08)$ \\
                     & $\Omegab h^2$       & Baryon density$~^b$      & $\mathcal{N}(0.0214, 0.002)$ \\
                     & $\Omegac h^2$       & Cold dark matter density & \\
                     & $\ln(10^{10} \mysub{A}{s})$ & Scalar power spectrum amplitude & \\
                     & $\mysub{n}{s}$               & Scalar power spectrum slope$~^{b,c}$ & $=0.95$ \\
                     & $\tau$              & Optical depth to reionization$~^c$ & \\
                     & $\mysub{A}{SZ}$     & Sunyaev-Zel'dovich signal amplitude$~^c$ & $\mathcal{U}(0, 2)$ \\
                     & $w$                 & Constant dark energy equation of state & \\
                     & $w_0$               & Evolving $w$: current value & \\
                     & $w_a,\wet$          & Evolving $w$: value at early times & \\
                     & $\mysub{a}{t}$      & Evolving $w$: transition scale factor & $\mathcal{U}(0.5, 0.95)$\vspace{1mm} \\
    Mass function (\S\ref{sec:massfunction})
                     & $A$                 & Global amplitude & $\mathcal{N}_4(0.20,\ldots)$ \\
                     & $a$                 & Low mass amplitude & $\mathcal{N}_4(1.52,\ldots)$ \\
                     & $b$                 & Low mass slope & $\mathcal{N}_4(2.25,\ldots)$ \\
                     & $c$                 & Exponential cutoff scale & $\mathcal{N}_4(1.27,\ldots)$ \\
                     & $\varepsilon$       & Evolution strength & $\mathcal{N}(1.0,0.1)$\vspace{1mm} \\
    Scaling relation (\S\ref{sec:scalingmodel})
                     & $\beta_0^{\ell m},\beta_1^{\ell m}$ & Nominal luminosity--mass relation & \\
                     & $\beta_0^{tm},\beta_1^{tm}$       & Nominal temperature--mass relation & \\
                     & $\sigma_{\ell m},\sigma_{tm}$ & Marginal scaling relation scatters & \\
                     & $\rho_{\ell t m}$     & Scaling relation scatter correlation & \vspace{1mm} \\
    Other (\S\ref{sec:samplingmodel}, \ref{sec:surveymodel}) 
                     & ---                 & \Chandra{} temperature calibration & 10\% Normal \\
                     & $\eta_g$            & Cluster gas mass  profile logarithmic slope & $\mathcal{N}(1.092,0.006)$ \\
                     & $\eta_L$            & Cluster luminosity profile logarithmic slope & $\mathcal{N}(0.1135,0.0005)$ \\
                     & $\kappa$            & $r_{500}$-to-survey flux conversion$~^d$ & \\
                     & $\xi$               & Completeness/purity$~^d$ & $\mathcal{U}(0.95,1.05)$ \\
    \hline
  \end{tabular}
  \label{tab:paramlist}
\end{table*}

\subsection{Cosmological models} \label{sec:cosmodel}

The simplest cosmological model considered in this study contains dark energy in the form of a spatially uniform and non-evolving energy density, i.e. a cosmological constant (\LCDM). In this model, the cosmological parameters relevant for the growth of structure are the mean baryon density, \Omegab{}; the mean total matter density, \Omegam{}; the Hubble parameter, $h$; and the matter power spectrum normalization, $\sigma_8$. Here the mean densities refer to redshift zero, since their values at other times are then determined by the Friedmann equation, and $\sigma_8^2$ is the $z=0$ variance in the density field at scales of $8h^{-1}\Mpc$, defined explicitly in \eqnref{eq:sigma2def} below. Because the mass range of our data corresponds to a small range in scale, we do not simultaneously fit for the spectral index of scalar density perturbations, $\mysub{n}{s}$, but rather fix its value at 0.95 \citep[e.g.][]{Komatsu09}, except when simultaneously fitting CMB data (see \secref{sec:data} and \tabref{tab:paramlist}). This assumption does not significantly affect our results; see \secref{sec:lcdmres}. We assume that the Universe is spatially flat on large scales throughout.

We additionally consider models in which dark energy is a fluid parametrized by a constant equation of state, $w$ (constant $w$ models). Unlike the cosmological constant scenario, such a fluid will not in general have uniform density, and thus contributes to some degree to the evolution of density perturbations, in addition to influencing the expansion history of the Universe. Due to theoretical uncertainties on the behavior of the dark energy fluid on the nonlinear scales that determine the mass function, numerical simulations of the mass function have to date been done only for models in which dark energy is uniform, even when $w \neq -1$. Our approach is to straightforwardly propagate the influence of non-uniform dark energy on linear scales but to leave the mass function unaltered, while continuing to use our standard systematic allowances on the mass function and its evolution (\secref{sec:massfunction}). We have verified that the value of the dark energy sound speed (assumed to be constant with time) has no effect on our results.  Preliminary theoretical work indicates that the effect of dark energy perturbations on the mass function might be readily measurable (\citealt*{Abramo09}; \citealt{Creminelli09,Hwang09,Alimi09}), in which case our approach likely underestimates the ability of the data to discriminate among these models.

Finally, we consider models in which the dark energy equation of state is a function of time, according to two parametrizations. The first is the commonly used model of \citet{Chevallier01} and \citet{Linder03a},
\begin{equation}
  \label{eq:lindermod}
  w(a) = w_0 + w_a(1-a),
\end{equation}
where $a=1/(1+z)$ is the scale factor, in which the equation of state makes a smooth transition from value $w_0$ at $z=0$ to $w_0+w_a$ at high redshift. A generalization due to \citet{Rapetti05},
\begin{equation}
  \label{eq:wevolmod}
  w(z) = \frac{\wet z + w_0 \mysub{z}{t}}{z + \mysub{z}{t}},
\end{equation}
has the advantage that the transition redshift, $\mysub{z}{t}$, can be marginalized over. (\eqnref{eq:lindermod} is a special case of \eqnref{eq:wevolmod} with $\mysub{z}{t}=1$ and $w_a=\wet-w_0$.) This model has greater applicability to current data, which primarily constrain $w$ at $z<1$, resulting in more commensurate constraints on the current and early-time equation of state, $w_0$ and $\wet$. In practice, we marginalize over the scale factor of the transition, $\mysub{a}{t}$, within the range $0.5 < \mysub{a}{t} < 0.95$.

\subsection{Mass function} \label{sec:massfunction}

Cosmological analyses of the kind presented here are enabled by the fact that, to good approximation, the expected number density of dark matter halos as a function of mass, $M$, can be expressed as a relatively simple function of cosmological parameters,
\begin{equation}
  \label{eq:massfunction}
  \frac{dn(M,z)}{dM} = \frac{\rhom}{M} \frac{d\ln \sigma^{-1}}{dM} f(\sigma).
\end{equation}
Here $\rhom$ is the mean comoving matter density and $\sigma^2$ is the variance of the linearly evolved density field, smoothed by a spherical top-hat window of comoving radius $r$, enclosing mass $M=4\pi \rhom r^3 / 3$,
\begin{equation}
  \label{eq:sigma2def}
  \sigma^2(M,z) = \frac{1}{2\pi^2} \int_0^\infty k^2 P(k,z) |W_M(k)|^2 dk,
\end{equation}
where $P(k,z)$ is the linear power spectrum evolved to redshift $z$ and $W_M(k)$ is the Fourier transform of the window function. In the formulation of \eqnref{eq:massfunction}, the mass function depends on cosmological parameters and redshift only through $\sigma^2(M,z)$. The function $f(\sigma)$ may be an analytic or semi-analytic approximation \citep{Press74,Bond91,Sheth99} or a fit to cosmological $N$-body simulations.

The applicability of this ``universal'' form of the mass function was first demonstrated in numerical dark matter simulations of flat \LCDM{} and open ($\Omegal=0$) cosmologies by \citet{Jenkins01} and confirmed by \citet{Evrard02}. It has since been verified that the fitting function provided by Jenkins is approximately accurate (within $\sim 20$ per cent) among models with constant $w\neq-1$ and some evolving $w$ models \citep{Klypin03,Linder03,Lokas04,Kuhlen05}.\footnote{We reiterate that these works include the effects of dark energy on the mass function through the cosmic expansion rate, but not the effects of dark energy density perturbations (\secref{sec:cosmodel}).} Other authors have studied the dependence of $f(\sigma)$ on redshift beyond that implicit in $\sigma^2(M,z)$ \citep{Lukic07,Reed07,Cohn08}, replacing $f(\sigma)$ with $f(\sigma,z)$. The most recent and relevant work is that of \citet{Tinker08}, which we adopt here.

The Tinker fitting function has the form
\begin{equation}
  \label{eq:tmf}
  f(\sigma,z) = A \left[ \left( \frac{\sigma}{b} \right)^{-a} + 1 \right] e^{-c/\sigma^2},
\end{equation}
where each of the fitted parameters has a redshift dependence of the form
\begin{equation}
  \label{eq:tmfz}
  x(z) = x_0(1+z)^{\varepsilon \alpha_x}; ~~ x \in \{A,a,b,c\}.
\end{equation}
The various parameters $x_0$ and $\alpha_x$ are given in \citet{Tinker08} as a function of the spherical overdensity, $\Delta$, used to define the cluster radius. Unlike the $z=0$ mass function, this additional redshift dependence has not been tested in simulations of cosmologies beyond the simple, flat \LCDM{} model. We therefore introduce the parameter $\varepsilon$, which controls the overall strength of the evolution given by the $\alpha_x$, in order to marginalize over remaining uncertainties in the redshift dependence of the mass function in exotic cosmologies. We also choose to work with the $\Delta=300\Omegam(z)$ fit to $f(\sigma,z)$ (a relatively large cluster radius) because the evolution parametrized by the $\alpha_x$ becomes more pronounced with increasing overdensity (smaller radius).

To address the uncertainty in the normalization and shape of $f(\sigma,z=0)$, we marginalized over each of the fitted parameters in \eqnref{eq:tmf} using the covariance matrix of the fit (Jeremy Tinker, private communication). The statistical error of this fit is $<5$ per cent; however, this figure does not reflect systematic uncertainties due to the presence of baryons \citep[e.g.][]{Stanek09}, evolving dark energy, etc. We therefore scaled the covariance matrix when defining this prior on the mass function parameters such that the marginal uncertainty at fixed $\log_{10}(\sigma^{-1})=0.2$ ($M\sim 10^{15} \Msun$ in the concordance model) is a conservative 10 per cent.\footnote{There is no reason, a priori, that the systematic uncertainty in the mass function should have a similar form to the statistical covariance. At minimum, however, this procedure provides a straightforward way to marginalize over a family of functions that are similar to the mass function, while still allowing differences in the shape as well as the normalization. As we show in \secref{sec:mfcnpriors}, uncertainty on the mass function at this level has negligible impact on our results, in any case.}

\subsection{Scaling relations} \label{sec:scalingmodel}

To perform the cosmological analysis, we need to relate cluster mass to the observable that determines cluster detection, in this case X-ray flux. Given a redshift, $z$, a cluster's unabsorbed, soft X-ray flux, $F$, is determined by its intrinsic X-ray luminosity, $L$, temperature, $kT$, and metallicity, $Z$, as
\begin{equation}
  \label{eq:Kcorrection}
  F(z,L,kT,Z) = \frac{L}{4 \pi d_L^{\hspace{0.2ex}2}(z) K(z,kT,Z)},
\end{equation}
where $d_L(z)$ is the luminosity distance to the cluster, and $K(z,kT,Z)$ is the required $K$-correction. For intracluster medium temperatures $kT>3\keV$ and luminosities and fluxes in the soft (\ROSAT{}) X-ray band (0.1--2.4\keV{}), $K$ has only a weak dependence on temperature and negligible dependence on metallicity; we hereafter fix $Z$ to the typical value of 0.3 times the solar value.

As discussed in \scalingpaper{}, a simple prescription for how $L$ and $kT$ are related to the total mass, $M$, is given by the self-similar model \citep{Kaiser86}. We define nominal luminosity--mass and temperature--mass relations at $r_{500}$ of the form
\begin{eqnarray}
  \label{eq:nominalMLT}
  \expectation{\ell(m)} &=& \beta_0^{\ell m} + \beta_1^{\ell m} m, \\
  \expectation{t(m)} &=& \beta_0^{tm} + \beta_1^{tm} m \nonumber
\end{eqnarray}
where
\begin{eqnarray}
  \label{eq:MLTdefs}
  \ell &=& \logTen\left(\frac{L_{500}}{E(z)10^{44}\erg\second^{-1}}\right), \nonumber\\
  m &=& \logTen\left(\frac{E(z)M_{500}}{10^{15}\Msun}\right), \\
  t &=& \logTen\left(\frac{kT_{500}}{\keV}\right), \nonumber
\end{eqnarray}
and $E(z)$ is the normalized Hubble parameter, $H(z)/H_0$. The factors of $E(z)$ appearing explicitly in \eqnref{eq:MLTdefs} follow from the definition of cluster radius using a fixed overdensity with respect to the critical density \citep{Bryan98}. To describe the intrinsic scatter in $\ell$ and $t$ given $m$ about the nominal relations, we adopt a simple, bivariate normal distribution parametrized by the marginal luminosity--mass and temperature--mass log-normal scatters, $\sigma_{\ell m}$ and $\sigma_{tm}$, and a coefficient of correlation, $\rho_{\ell t m}$.

There are various ways of adding complexity to this scaling relation model, including departures from self-similar evolution in the normalization of the nominal relations, evolution in the scatter, and asymmetry in the scatter. In \scalingpaper{}, we show that the data are consistent with the simple model defined above, and do not require or prefer any such additions, even when the cosmological parameters are extremely restricted by external data. We therefore adopt the simple model above in this work.

Motivated by the results of \citet{Evrard08}, we have defined the scaling relations for quantities within $r_{500}$, which also corresponds to our measurements of mass, luminosity and temperature from follow-up observations (\scalingpaper{}). To convert the mass definition used by the mass function, $\Delta=300\Omegam(z)$, to $\Delta=500$, we use the procedure of \citet{Hu03}, assuming a \citet*[][hereafter NFW]{Navarro97} mass distribution with concentration parameter $c=4$. This conversion is negligibly sensitive to the assumed concentration parameter, since both $r_{500}$ and $r_{300\Omegam(z)}$ are well beyond the NFW scale radius for reasonable values \citep[$c>3$;][]{Zhao03,Zhao08,Gao08}.

\subsection{Sampling model: follow-up observations} \label{sec:samplingmodel}

The next component of the model connects quantities predicted from the cosmology, mass function and scaling relations to quantities measured from the follow-up X-ray observations. Because some of our measurements are made with respect to a reference cosmology (see \scalingpaper{}), this procedure is not entirely trivial.

In particular, the mass, luminosity and temperature determined from the follow-up X-ray observations are measured within $r_{500}$, itself determined via the implicit equation
\begin{equation}
  \label{eq:rDelta}
  M(r_{500}) = \frac{\Mgas(r_{500})}{\fgas(r_{500})} = \frac{4\pi}{3}(500)\rhoc(z) r_{500}^3,
\end{equation}
which can be re-written
\begin{equation}
  \Mgas(r) \propto \rhoc(z) r^3 \fgas(r) \propto r^{\eta_g},
\end{equation}
where $\eta_g$ is the logarithmic slope of the gas mass profile at large radius. Using that $\rhoc(z) \propto H^2(z)$, the expression
\begin{equation}
  \label{eq:rscaling}
  r_{500} \propto \left[ \fgas(r_{500}) H^2(z) \right]^{1/(\eta_g-3)}
\end{equation}
relates the ``true'' value of $r_{500}$ predicted by a set of model parameters to the value of $r_{500}$ that we would have inferred assuming our reference cosmology and reference \fgas{} value. The gas mass profiles measured in \scalingpaper{} are self similar, consistent with a constant value of $\eta_g$; for simplicity, we therefore adopt $\eta_g=1.092\pm 0.006$, determined from a fit to the entire sample from 0.7--1.3$r_{500}$, and marginalize over the uncertainty.

To see how the measurements of total cluster mass depend on model parameters, we first write
\begin{eqnarray}
  \frac{\mysup{M}{ref}(r)}{M(r)} &=& \frac{\mysup{\Mgas}{ref}(r)/\mysup{\fgas}{ref}(r)}{\Mgas(r)/\fgas(r)} \mysub{R}{NFW} \\
  &=& \frac{\mysup{d_A}{ref}(z)^{2.5}\fgas}{d_A(z)^{2.5}\mysup{\fgas}{ref}} \mysub{R}{NFW}, \nonumber
\end{eqnarray}
where $\mysup{M}{ref}$ is the prediction for what mass would be measured for a cluster of true mass $M$ using our assumed reference parameter values, and $d_A(z)$ is the angular diameter distance to redshift $z$. $\mysup{d_A}{ref}(z)$ and $\mysup{\fgas}{ref}$ are not predictions for measured values, but depend directly on the reference cosmology and \fgas{} value. The first term in this expression accounts for the dependence of the mass measured within a fixed angular aperture on distance and \fgas{}, given that the mass is estimated via the gas mass and gas mass fraction, and where we have used the scaling $\Mgas\propto d_A(z)^{2.5}$. The dependence on the angular size corresponding to physical radius $r$ is handled by $\mysub{R}{NFW}$, which we evaluate assuming that the shape of the total mass distribution near $r_{500}$ is well approximated by the NFW profile; in this case the scaling factor is straightforward to compute using the scaling of $r_{500}$ given in \eqnref{eq:rscaling}:
\begin{equation}
  \label{eq:RNFW}
  \mysub{R}{NFW} = \frac{\ln(1+x c_{500}) - x c_{500} / (1 + x c_{500})}{\ln(1+c_{500}) - c_{500} / (1 + c_{500})}
\end{equation}
where $x=\mysup{r_{500}}{ref}/r_{500}$. To evaluate this factor, we assume a concentration parameter $c=4$,\footnote{Note that, by convention, $c$ without a subscript refers to $c_{200}$, defined as $r_{200}$ in units of the NFW scale radius; $c=4$ corresponds to $c_{500} \approx 2.6$.} although we note that $\mysub{R}{NFW}$ is extremely insensitive to this assumption provided that $c > 3$ (i.e. provided $r_{500}$ is well beyond the scale radius).

Similarly, we can write the scaling of the luminosity as
\begin{equation}
  L_{500}(r) \propto d_L^{\hspace{0.2ex}2}(z) \left( \frac{r_{500}}{d_A(z)} \right)^{\eta_L},
\end{equation}
where $d_L(z)$ is the luminosity distance and the first term is the scaling of the luminosity measured within a fixed angle; the second term is simply the angular size of $r_{500}$ raised to the power of $\eta_L$, the logarithmic slope of the integrated luminosity profile. As with the gas mass profiles, we fit this slope between 0.7 and 1.3$r_{500}$, finding $\eta_L=0.1135\pm 0.0005$ from the entire sample, and marginalize over the uncertainty.

Since the emission-weighted, average temperature within $r_{500}$ is a weak function of $r_{500}$ compared with a typical statistical error bar, and considering the relative insensitivity of the cosmological analysis to the precise $kT$ values, we do not similarly model any dependence of the measured temperatures on model parameters. However, we marginalize over a global 10 per cent Gaussian systematic uncertainty on the temperature measurements to account for residual uncertainty in instrument calibration.

Since the masses, luminosities and temperatures from the follow-up observations are determined from the same data, their statistical error bars are correlated. The masses and luminosities are integrated within $r_{500}$, so their errors are primarily correlated via the uncertainty in $r_{500}$. The luminosity and temperature errors are also somewhat correlated, since the K-corrections that convert flux to luminosity have a weak temperature dependence. (There is no correlation between mass and temperature errors using our method; see \scalingpaper{}.) Since these measurement errors were propagated via Monte Carlo, it is straightforward to compute the error correlations from the samples; we find $\rho_{\hat{m}\hat{\ell}}=0.29 \pm 0.19$ and $\rho_{\hat{\ell}\hat{t}}=-0.4 \pm 0.3$, averaged over the cluster sample. Note that the latter quantity is not related to $\rho_{\ell t m}$, defined in \secref{sec:scalingmodel}, which quantifies the correlation between departures in luminosity and temperature from the nominal scaling relation, not the correlation of measurement errors.

\subsection{Sampling model: survey} \label{sec:surveymodel}

The final ingredient is a sampling model for the selection of our clusters from the RASS, incorporating the selection function of each cluster sample, the survey flux measurement errors, and the relation between the flux measurements reported in each cluster sample and $F_{500}$, the flux within $r_{500}$ from follow-up observations (in the same energy band). The fluxes reported for each cluster sample are measured differently \citep[see][]{Ebeling98,Ebeling01,Bohringer04} and none corresponds to a flux measurement within a fixed-overdensity radius such as $r_{500}$. Fortunately, we find a good empirical correlation between the reported survey fluxes and our fluxes measured from the follow-up observations for clusters well above the flux limit.\footnote{For true fluxes near or below the flux limit, Malmquist bias distorts the relation.} We include a simple linear conversion between survey flux, $F_s$, and follow-up observation flux, $F_{500}$, in the model,
\begin{equation}
  F_s = \kappa F_{500},
\end{equation}
where the slope, $\kappa$, is allowed to be different for each cluster sample. Because the survey fluxes suffer from Malmquist and Eddington biases, the $\kappa$'s must be constrained simultaneously with the full model, including the cosmology and scaling relation. These parameters also trivially account for any residual uncertainties in the flux cross-calibration between \ROSAT{} and \Chandra{} (see \scalingpaper{}).

The flux measurement error as a function of reported flux for each sample is consistent with Poisson scaling, 
\begin{equation} \label{eq:Fserr}
  \sigma_F\propto \sqrt{F_s},
\end{equation}
although the normalization varies by sample due to the different algorithms used to measure the flux. We approximate the survey flux sampling distribution as Gaussian, centered on the true flux, with width determined via \eqnref{eq:Fserr}, fitting a separate normalization for each cluster sample directly from the reported fluxes and errors.

Finally, the selection function of each cluster sample, the probability that a cluster is included in the sample as a function of redshift and measured survey flux, is modeled by interpolating the look-up tables provided for each sample (\citealt{Ebeling98}, 2010; \citealt{Bohringer04}). Conservatively, we include an additional nuisance parameter to marginalize over possible uniform incompleteness or impurity at the 5 per cent level,
\begin{equation}
  \mysub{P}{sel}(z,\hat{F}_s) \propto \xi,
\end{equation}
where each sample can have a different value of $\xi$.

\section{Analysis Method} \label{sec:method}

\subsection{Likelihood} \label{sec:likelihood}

Early cosmological work using galaxy cluster samples used an approach based around binning the detected clusters in redshift and flux (or luminosity) and either adopting external priors on the luminosity--mass relation \citep[e.g.][]{Borgani01} or simultaneously fitting an external luminosity--mass data set without accounting for selection effects \citep[e.g.][]{Allen03}.\footnote{We note that some authors have followed the (in principle) equivalent approach of using a scaling relation to transform the predicted mass function into, e.g. the baryonic mass function \citep{Voevodkin04} or X-ray temperature function \citep{Henry04,Henry09}. Although the statistical formalism discussed in this section is considered as an extension of  the flux-redshift binning approach, it also underlies these methods.} In \citetalias{Mantz08}, we took this approach to its logical limit by deriving the likelihood for bins of infinitesimal volume; however, that work still suffered from the fact that it used an external data set to constrain the luminosity--mass relation without explicitly accounting for selection bias. Consequently, it was necessary to restrict that external data set \citep{Reiprich02} to low redshifts and high fluxes in order to minimize the effects of selection bias, making it impossible to test for departures from self-similar evolution in the scaling relation. More recently, \citet{Vikhlinin09,Vikhlinin09a} binned their detected clusters  in redshift and mass (again with infinitesimally small bins) and used the same cluster sample to constrain the scaling relations; however, their procedure still does not produce a self-consistent fit for both scaling relations and cosmology.

In this section, we show that our procedure from \citetalias{Mantz08} can be generalized to allow such a simultaneous and self-consistent fit, using follow-up observations of flux-selected clusters to constrain the scaling relations over the full redshift range of the data, and accounting fully for the presence of Malmquist and Eddington biases. We also show that the corresponding likelihood function can be derived from first principles, beginning with a simple Bayesian regression model. For simplicity, we derive the likelihood for the general problem of counting sources as a function of their properties. This general picture includes the following components:
\begin{enumerate}
\item A population function,\footnote{Note that this nomenclature is not widely used. We adopt the term ``population function'' to imply an analogy to the mass function used in this study.} $\expectation{dN/dx}$, which provides a theoretical prediction for the underlying distribution of sources (i.e. their number, $N$) as a function of properties, $x$ (see below).
\item Population variables, $x$, on which the population function depends.
\item Response variables, $y$, which obey a stochastic scaling relation as a function of $x$.
\item The stochastic scaling relation, $P(y|x)$.
\item Observed values, $\hat{x}$ and $\hat{y}$. Note that all of the $x$ and $y$ need not actually be measured, the exception being those which determine whether a source is included in the sample (detected; see also \secref{sec:appgos}).
\item Sampling distributions for the observations as a function of population and response variables, $P(\hat{x},\hat{y}|x,y)$.
\item A selection function, $P(I|x,y,\hat{x},\hat{y})$. $I$ represents the inclusion of a source in the sample. In full generality, the selection function could be a function of true variables as well as observations; for example, if it is calibrated using reliable simulations.
\end{enumerate}
Throughout the following, we make the approximation that the clustering of sources is unimportant compared with the pure Poisson nature of their occurrence, which is justified for current all-sky surveys of very massive clusters (i.e. where the survey dimensions are large compared with the correlation length), as in this study \citep{Hu03,Holder06}. Provided this assumption remains valid, the approach described here can be applied equally well to cluster studies based on optical or Sunyaev-Zel'dovich (SZ) surveys. We note that this is only a simplifying assumption; in principle, the approach detailed below can be straightforwardly generalized to include the spatial correlation of sources.

\subsubsection{Binning derivation}
Divide the space $(\hat{x},\hat{y})$ into bins (indexed by $j$) of very small volume $\Delta\hat{x}_j\Delta\hat{y}_j$, such that no bin contains more than one detected source, and both the population function and scaling relation are approximately constant over the bin volume. The expected number of detected sources in bin $j$, located at $(\hat{x}_j,\hat{y}_j)$, is
\begin{eqnarray} \label{eq:growth_like_binN}
  \expectation{\Nobs{}_{,j}} & = & (\Delta\hat{x}_j\Delta\hat{y}_j) \int dx \int dy \expectation{\frac{dN}{dx}} P(y|x) \nonumber\\
  & & \times ~ P(\hat{x}_j,\hat{y}_j|x,y) P(I|x,y,\hat{x}_j,\hat{y}_j).
\end{eqnarray}

The likelihood, \like{}, is a product of independent Poisson likelihoods for each bin,
\begin{eqnarray} \label{eq:growth_like_binlike}
  \like(\{N_j\}) & = & \prod_j \frac{\expectation{\Nobs{}_{,j}}^{N_j} e^{-\expectation{\Nobs{}_{,j}}}}{N_j!} \nonumber \\
  & = & e^{-\expectation{\Nobs}} \prod_{j:N_j=1} \expectation{\Nobs{}_{,j}},
\end{eqnarray}
where $N_j$ is the actual number of clusters detected in bin $j$, $\expectation{\Nobs}$ is the predicted total number of detected sources (summed over all bins), and the second equality follows from the fact that $N_j \in \{0,1\}$ by construction.

Since the binning scheme must be invariant under changes to the values of the model parameters, the derivative of the log-likelihood is independent of the bin volumes, $\Delta\hat{x}_j\Delta\hat{y}_j$. Defining $\expectation{\nobs{}_{,j}} = \expectation{\Nobs{}_{,j}}/(\Delta\hat{x}_j\Delta\hat{y}_j)$, we can write
\begin{equation}
  \label{eq:growth_like_propto}
  \like(\{N_j\}) \propto e^{-\expectation{\Nobs}} \prod_{j:N_j=1} \expectation{\nobs{}_{,j}}
\end{equation}
and ignore the constant of proportionality for the purposes of maximizing \like.

\vspace{-2mm}
\subsubsection{Regression derivation}
In the regression of truncated data (where some sources are not detected and the total number of sources is thus unknown), the total number of detected plus undetected sources, $N$, becomes a parameter of the model and must be marginalized over \citep[for background, see][]{Gelman04,Kelly07}. We define $\expectation{N}$, $\expectation{\Nobs}$ and $\expectation{\Nmis}$ to be respectively the predictions for the total number of sources, number of detected sources, and number of undetected (missed) sources as follows:
\begin{eqnarray} \label{eq:bayesregdef}
  \expectation{N} & = & \int dx \expectation{\frac{dN}{dx}}, \nonumber \\
  \expectation{\Nobs} & = & \int dx \expectation{\frac{dN}{dx}} \int dy ~P(y|x) \\
  & & \times ~ \int d\hat{x} \int d\hat{y} ~P(\hat{x},\hat{y}|x,y) P(I|x,y,\hat{x},\hat{y}), \nonumber \\
  \expectation{\Nmis} & = & \expectation{N} - \expectation{\Nobs}. \nonumber
\end{eqnarray}

The joint likelihood of the observations and $N$ is
\begin{eqnarray} \label{eq:growth_like_bayeslike}
  \like(\hat{x},\hat{y},N) & = & \left[ \frac{\expectation{N}^N e^{-\expectation{N}}}{N!} \right]
  \left[ \frac{N!}{\Nobs!\Nmis!} \right] \\
  & & \times \prod_{i=1}^{\Nobs} \mysub{P}{det}(\hat{x}_i,\hat{y}_i,I) \prod_{j=1}^{\Nmis} \mysub{P}{mis}\left(\bar{I}\right). \nonumber
\end{eqnarray}
The first term is a Poisson likelihood for the model parameter $N$, the total number of sources. The second factor is a binomial coefficient enumerating the number of ways of selecting $\Nobs$ detected sources from $N$. This term is necessary because, in the Bayesian treatment, the true source properties $(x,y)$ are random variables; as such, sources are distinguishable only in terms of their detection ($I$) or non-detection ($\bar{I}$). The probability that the \Nobs{} detected sources have measurements $(\hat{x},\hat{y})$ is accounted for by the product over detected objects. The second product accounts for the likelihood of not detecting an additional \Nmis{} sources whose properties are distributed according to the population function and scaling relation.

Noting that the marginal probability for a source to have properties $x$ is
\begin{equation}
  P(x) = \frac{\expectation{dN/dx}}{\expectation{N}},
\end{equation}
the functions \mysub{P}{det} and \mysub{P}{mis} can be related to quantities defined previously (cf. \eqnrefs{eq:growth_like_binN} and \ref{eq:bayesregdef}):
\begin{eqnarray} \label{eq:bayes_Pobs}
  \mysub{P}{det}(\hat{x}_i,\hat{y}_i,I) & = & \int dx \int dy \frac{\expectation{dN/dx}}{\expectation{N}} P(y|x) \nonumber \\
  & & \times ~ P(\hat{x}_i,\hat{y}_i|x,y) P(I|x,y,\hat{x}_i,\hat{y}_i) \nonumber \\
  & = & \frac{\expectation{\nobs{}_{,i}}}{\expectation{N}}
\end{eqnarray}
and
\begin{eqnarray} \label{eq:bayes_Pmis}
  \mysub{P}{mis}\left(\bar{I}\right) & = & \int dx \int dy \frac{\expectation{dN/dx}}{\expectation{N}} P(y|x) \nonumber \\
  & & \times ~ \int d\hat{x} \int d\hat{y} ~P(\hat{x},\hat{y}|x,y) P(\bar{I}|x,y,\hat{x},\hat{y}) \nonumber \\
  & = & \frac{\expectation{\Nmis}}{\expectation{N}}.
\end{eqnarray}
Substituting these expressions, the likelihood simplifies to
\begin{eqnarray} \label{eq:growth_like_bayesL}
  \like(\hat{x},\hat{y},N) & = & \left[ \frac{\expectation{N}^N}{\expectation{N}^{\Nobs}\expectation{N}^{\Nmis}} \right] \nonumber
  \left[ \frac{1}{\Nobs!} \right] \\
  & & \times ~ \left[ \frac{\expectation{\Nmis}^{\Nmis} e^{-\expectation{\Nmis}}}{\Nmis!} \right] \nonumber \\
  & & \times ~ e^{-\expectation{\Nobs}} \prod_{i=1}^{\Nobs} \expectation{\nobs{}_{,i}}.
\end{eqnarray}
The first factor above is unity, and the second is a normalization term independent of model predictions. The third term is a Poisson density which sums to unity when \like{} is marginalized over $N$ from \Nobs{} to infinity (equivalently, \Nmis{} from zero to infinity). The remaining terms are exactly equivalent to those in \eqnref{eq:growth_like_propto}.\footnote{The normalization factors present in \eqnrefs{eq:growth_like_binlike} and \ref{eq:growth_like_bayesL} are different because the two likelihood functions are formally defined over different domains. What is significant is that both functions have the same dependence on model predictions.}

\subsubsection{Application to the growth of structure} \label{sec:appgos}
The formalism of the last sections can be straightforwardly applied to observations of the growth of cosmic structure. The population variables describing clusters are redshift, $z$, and mass, $m$, and the population function is simply the product of the mass function and the comoving volume element, $dN/dzdm=(dn/dm)(dV/dz)$ (see definitions in \secref{sec:modeling}). The response variables are luminosity, $\ell$, and temperature, $t$, which scale with mass and redshift as described in \secref{sec:scalingmodel}. The observations include spectroscopically determined redshifts (for which we neglect the measurement errors in practice), the masses, luminosities and temperatures determined from follow-up observations, and the survey fluxes. Their sampling distributions, which account for the correlations among $m$, $\ell$ and $t$ measurements from the same follow-up data, are discussed in \secrefs{sec:samplingmodel} and \ref{sec:surveymodel}. The selection functions of the cluster samples are given as functions of redshift and observed survey flux, and incorporate the appropriate sky coverage fraction.

It is important to note that this analysis method does not require all the potentially observed values $\{\hat{x},\hat{y}\}$ to be available for each source, apart from those which determine whether a source is included in the sample.\footnote{However, the observations do have to span the range of values present in the sample reasonably well, in order to adequately probe the scaling relations.} This is perhaps most apparent from considering how \eqnref{eq:bayes_Pobs} would be modified if a particular observable, $\hat{y}_0$, were unavailable: an additional integration over $\hat{y}_0$ would be required, which would result in $P(\hat{y}_0|x,y)$ being replaced by unity.\footnote{This simplification occurs because the selection function is necessarily not dependent on $\hat{y}_0$. Similarly, in practice, \eqnref{eq:bayes_Pmis} need be integrated only over observables that influence selection.} In the binning derivation, this is analogous to the bin containing that cluster being unbounded rather than constrained in the $\hat{y}_0$ direction. This result is intuitive, as it simply means that the integral over $y_0$ for that cluster is not weighted by the information gained from a measurement.

Thus, our analysis makes use of the available information for all 238 clusters meeting our selection criteria. For the 144 clusters without follow-up data, only the redshift and survey fluxes are available; following the argument above, the terms associated with follow-up luminosity, mass and temperature measurements for each of these clusters integrate to unity. In effect, the likelihood for these clusters (here we refer to the terms $\expectation{\nobs{}_{,j}}$) simplify to the form used in \citetalias{Mantz08}, which depend only on observations of redshift and survey flux. For the 94 clusters with follow-up measurements, the analysis uses the redshift and survey flux, in addition to the (independently measured) follow-up luminosity, mass and temperature. The latter are, of course, crucial for obtaining simultaneous constraints on the scaling relations.

\begin{figure*}
  \centering
  \includegraphics[scale=\figscl]{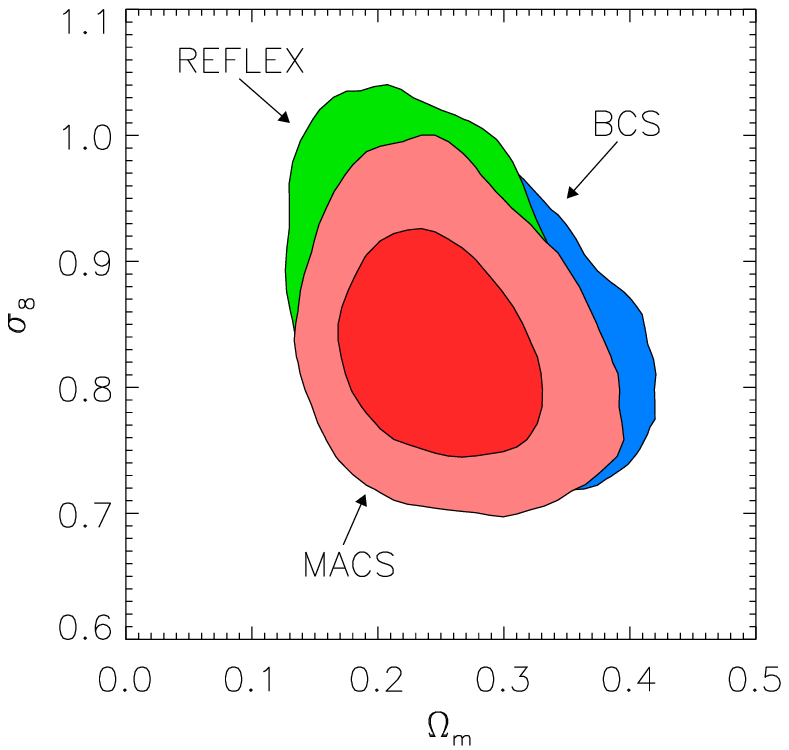}
  \hspace{1cm}
  \includegraphics[scale=\figscl]{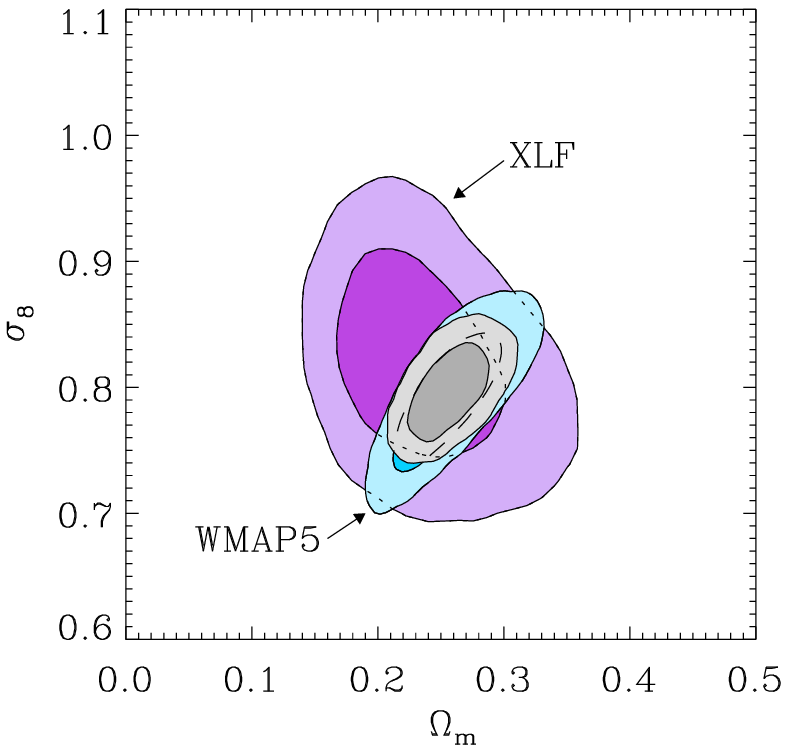}
  \caption{Left: Joint 68.3 and 95.4 per cent confidence regions for parameters of the \LCDM{} cosmology from the BCS (blue), REFLEX (green) and Bright MACS (red) cluster samples individually, including all systematic allowances in \tabref{tab:paramlist}. Note that only the 95.4 per cent confidence regions are visible for BCS and REFLEX. Right: Constraints from the full XLF data set (purple) and 5-year WMAP data \citep[blue;][marginalized over the SZ signal]{Dunkley09}. Results from the combination of the XLF and WMAP5 data are shown in gray.}
  \label{fig:LCDMfig}
\end{figure*}

The ability to use partial follow-up in a statistically rigorous way has significance for the planning of large, future cluster surveys. In particular, it reveals as overly simplistic the choice between strategies employing exhaustive follow-up observations to calibrate scaling relations over a relatively small mass range and those using ``self-calibration'', which use many more sources but rely exclusively on the shape of the mass function to provide information about the scaling relations. Indeed, obtaining follow-up data for a fair sub-sample of the detected clusters potentially allows robust calibration of the scaling relations over a wide mass range, without unnecessarily limiting the size of the sample used for cosmology.\footnote{\citet{Sahlen09} demonstrate the improvement in cosmological constraints available to future surveys by incorporating follow-up observations.} However, we note that this attractively simple strategy is not necessarily the best; a more complete analysis, including an optimization procedure similar to that employed by \cite*{Wu09}, is warranted in planning the follow-up of future surveys.

Finally, we notice that the procedure described above is a conceptually simple generalization of that of \citetalias{Mantz08}. In fact, \eqnref{eq:growth_like_propto} is identical to Equation~18 in \citetalias{Mantz08}, with the exception that our bins are now defined in the higher-dimensional space including all observable quantities, rather than only survey flux and redshift.

\subsection{Mechanics and priors}

Using the likelihood function detailed in \secref{sec:likelihood}, cosmological constraints were obtained via Markov Chain Monte Carlo (MCMC), employing the Metropolis sampler embedded in the {\sc cosmomc} code\footnote{\url{http://cosmologist.info/cosmomc/}} of \citet{Lewis02}. The May 2008 {\sc cosmomc} release includes the 5-year \WMAP{} and Union supernova data and analysis codes; an additional module implementing the \fgas{} analysis has also been publicly released \citep[][\citetalias{Allen08}]{Rapetti05}.\footnote{\url{http://www.stanford.edu/~drapetti/fgas\_module/}} Further modifications were made to include the likelihood codes for the XLF and BAO data. The CMB and matter power spectrum calculations were performed using the {\sc camb} package of \citet*{Lewis00},\footnote{\url{http://www.camb.info}} suitably modified to incorporate the evolving $w$ models of \secref{sec:cosmodel} \citep{Rapetti05}.

The set of parameters describing the complete model is summarized in \tabref{tab:paramlist}, along with the priors used in our analysis.

\subsection{Improvements over M08} \label{sec:improvements}

Compared with our previous work in \citetalias{Mantz08}, the constraints on flat \LCDM{} and constant $w$ models reported below are tighter by a factor of 2--3. Given that we have used essentially the same cluster samples in the two studies, it is sensible to address the specific differences in the two analyses that result in these improvements. We attribute the change to several closely related improvements in our analysis.
\begin{enumerate}
\item Cluster mass determination: In \citetalias{Mantz08}, we calibrated the X-ray luminosity--mass relation using the HIFLUGCS data of \citet{Reiprich02}, for which masses were estimated from X-ray data, assuming hydrostatic equilibrium. This procedure is known to underestimate the mass by tens of per cent \citep[e.g.][]{Nagai07,Mahdavi08}, since the intracluster medium is not supported solely by thermal pressure. However, the overall size of this effect is not well established, nor is the variation in non-thermal pressure from cluster to cluster. Consequently, that earlier work was forced to incorporate a large systematic uncertainty in the mass measurements, which significantly degraded constraints on the scaling relation and its intrinsic scatter, resulting in a corresponding degradation in cosmological constraints.

The present work avoids this issue by using gas mass as a proxy for total mass (see details in \scalingpaper{}). Unlike total mass, gas mass can be measured from X-ray data with very little bias. The relation between gas mass and total mass is provided by \citetalias{Allen08}; ultimately, the cluster mass scale is set using the hydrostatic method for the clusters in that work, i.e. the class of hot, dynamically relaxed clusters for which the bias due to non-thermal pressure is minimal.

\item Follow-up data over a range of redshifts: The statistical methods employed in \citetalias{Mantz08} lack the internal consistency of our new procedure (\secref{sec:likelihood}), making it impossible to rigorously incorporate follow-up data at redshifts much greater than zero.\footnote{In detail, the follow-up data were restricted to low redshifts ($z<0.11$) where the HIFLUGCS discovery space was large, i.e. where many clusters were found well above the flux limit. This procedure reduces the biasing effects of flux selection, although it does not eliminate them.} As a result, we could not directly constrain the evolution of the scaling relations, and instead marginalized over a range of possibilities, using conservative priors. With our new method, follow-up data spanning the full redshift range of cluster detections can be included, allowing evolution in the scaling relations to be tested directly (\scalingpaper{}). Relatedly, the distribution of follow-up data over a range in redshift improves the constraints on the dark energy equation of state.

\item Use of \fgas{} data: As discussed by \citet{Vikhlinin09a}, the method used to estimate masses has an effect on the constraints obtained because different mass proxies have different dependencies on distance, with gas mass ($\Mgas \propto d^{\,2.5}$) being more sensitive than temperature or $Y_X$. Our results are therefore sensitive to this choice at some level.

Another way to look at this is to notice that the \citetalias{Allen08} data that we use to constrain \fgas{} inherently contain information about \Omegam{} in addition (given a bound on the baryonic depletion of clusters). Thus, our results on \Omegam{} are primarily, though not entirely, driven by these \fgas{} data. With the value of \fgas{} constrained, the XLF data then determine $\sigma_8$ and $w$. Note that, as mentioned in \secref{sec:data}, we use only six clusters from \citetalias{Allen08} at redshifts $z<0.15$; due to this redshift restriction, these \fgas{} data do not produce a constraint on $w$ by themselves.
\end{enumerate}

Although it is not listed explicitly above, we emphasize that the new statistical method outlined in this section is ultimately the most important advantage over \citetalias{Mantz08}, since the improvements in follow-up data and mass determination cannot be fully or fairly exploited without it.

\section{Results} \label{sec:results}

\begin{figure}
  \centering
  \includegraphics[scale=\figscl]{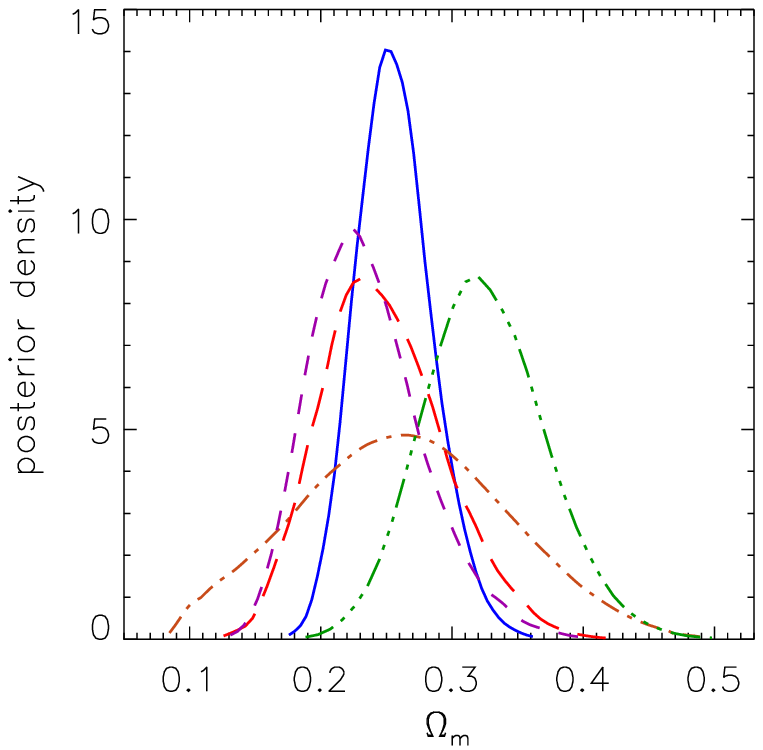}
  \caption{Marginalized posterior distributions for \Omegam{} in the \LCDM{} model from analysis of the XLF data (purple, short dashed line; including systematic allowances in \tabref{tab:paramlist}), 5-year WMAP data \citep[blue, solid line;][marginalized over the SZ signal]{Dunkley09}, cluster \fgas{} \citepalias[red, long dashed line;][including conservative systematic allowances]{Allen08}, SNIa \citep[green, 3-dot-dashed line;][including their treatment of systematics]{Kowalski08}, and BAO \citep[brown, dot-dashed line;][also using our standard priors on $h$ and $\Omegab h^2$]{Percival07}. Note that the XLF and \fgas{} results are not independent (\secrefs{sec:data} and \ref{sec:improvements}).}
  \label{fig:omegam}
\end{figure}

\subsection{Constraints on the \LCDM{} model} \label{sec:lcdmres}

\begin{figure*}
  \centering
  \includegraphics[scale=\figscl]{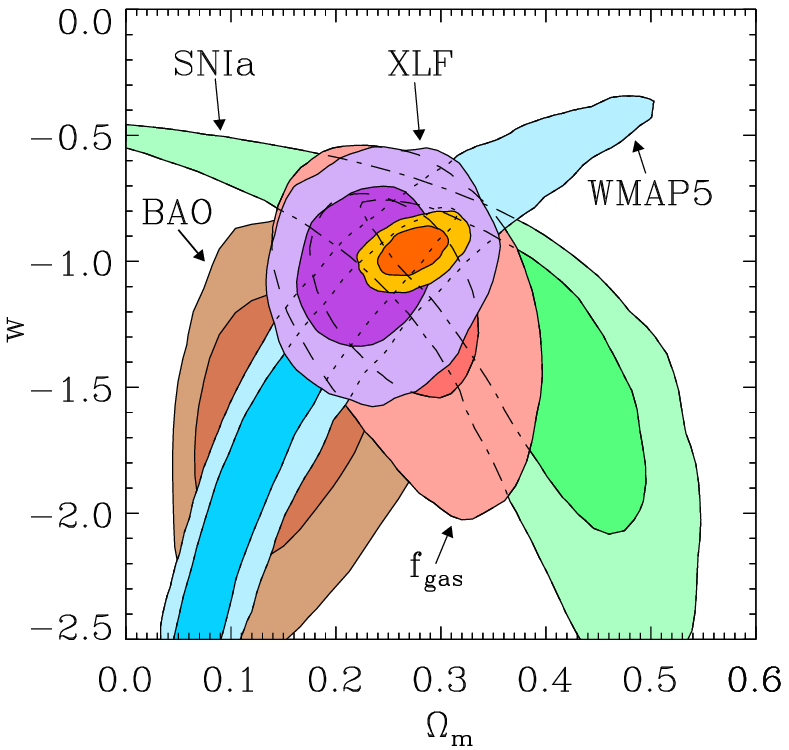}
  \hspace{1cm}
  \includegraphics[scale=\figscl]{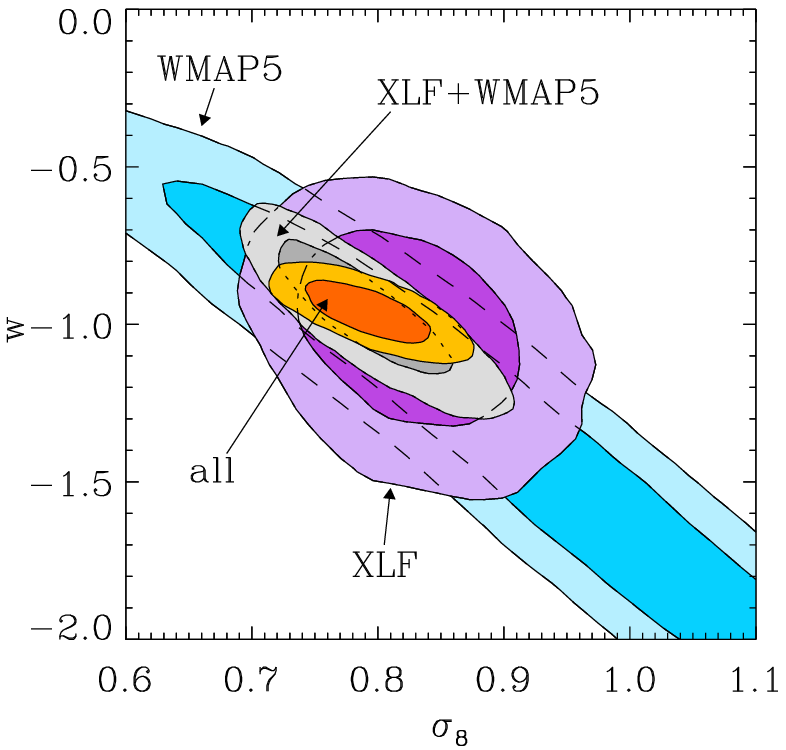}
  \caption{Joint 68.3 and 95.4 per cent confidence regions for parameters of the constant $w$ model. Left: constraints on \Omegam{} and $w$ from the XLF (purple, including all systematic allowances in \tabref{tab:paramlist}) are compared with those from cluster \fgas{} data \citepalias[red;][including conservative systematic allowances]{Allen08}, 5-year WMAP data \citep[blue;][marginalized over the SZ signal]{Dunkley09}, SNIa data \citep[green;][including their treatment of systematics]{Kowalski08}, and BAO observations \citep[brown;][also using our standard priors on $h$ and $\Omegab h^2$]{Percival07}. Results from combining these 5 data sets are shown in gold. Right: Constraints on $\sigma_8$ and $w$ from XLF and WMAP5 data. The combination of the XLF and WMAP5 yields the gray contours; adding the other data listed above produces the gold contours.}
  \label{fig:constantw}
\end{figure*}

For a spatially flat, cosmological constant ($w=-1$) model, the joint constraints on \Omegam{} and $\sigma_8$ obtained from the BCS, REFLEX, and Bright MACS cluster samples individually are displayed in the left panel of \figref{fig:LCDMfig}. Results from combining the 3 samples appear as purple contours in the right panel; the constraints, marginalized over the systematic allowances listed in \tabref{tab:paramlist}, are $\Omegam=0.23 \pm 0.04$ and $\sigma_8=0.82 \pm 0.05$ (\tabref{tab:results}). These results agree well with the tight constraints obtained from WMAP5 data \citep{Dunkley09} for this model (blue contours in the right panel). Results from  combination of WMAP5 and the XLF (gray contours) are somewhat improved: $\Omegam=0.26 \pm 0.02$ and $\sigma_8=0.80 \pm 0.02$. Our XLF results are in agreement with recent estimates based on other X-ray selected \citep{Henry09,Vikhlinin09a} and optically selected \citep{Rozo10} cluster samples, and a variety of independent cosmological data (\citealt{Percival07,Percival09}; \citetalias{Allen08}; \citealt[][see also references in \citetalias{Mantz08}]{Fu08,Ho08,Kowalski08,Hicken09,Reid09}).

As mentioned in \secref{sec:cosmodel}, the scalar spectral index, $\mysub{n}{s}$, is fixed at 0.95 in our analysis of the XLF data alone. The only parameter that is significantly degenerate with $\mysub{n}{s}$ is $\sigma_8$; our results for values of $\mysub{n}{s}$ other than 0.95 can be adequately described by shifting the $\sigma_8$ constraints along the linear relation $\sigma_8=0.82+0.25(\mysub{n}{s}-0.95)$.

The marginalized posterior distribution for \Omegam{} obtained from the XLF data is compared with those obtained from the analysis of WMAP5, \fgas{} \citepalias{Allen08}, SNIa \citep{Kowalski08}, and BAO \citep[][also using our standard priors on $h$ and $\Omegab h^2$]{Percival07} data in \figref{fig:omegam}. The agreement among all the data sets is good, though note that the XLF and \fgas{} results are not independent (\secrefs{sec:data} and \ref{sec:improvements}).

In \scalingpaper{}, we show that this simple cosmological model (and the simple scaling relation model of \secref{sec:scalingmodel}) provides an acceptable fit to the XLF data. Constraints on the parameters describing the scaling relations are also presented in that work.

\subsection{Constraints on constant $w$ models}

\begin{figure}
  \centering
  \includegraphics[scale=\figscl]{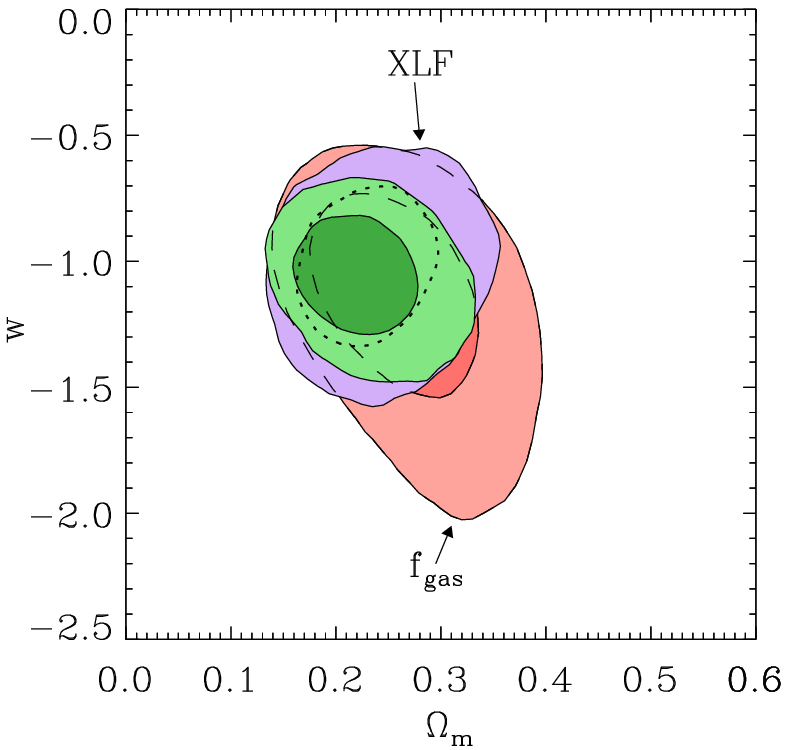}
  \caption{Joint 68.3 and 95.4 per cent confidence regions for parameters of the constant $w$ model, including conservative systematic allowances. Purple contours indicate constraints from the XLF (which includes six $z<0.15$ \fgas{} clusters; see \secrefs{sec:data} and \ref{sec:improvements}), while constraints from all 42 \fgas{} clusters alone are shown in red. Results combining the XLF data with all 42 \fgas{} clusters appear in green. The combination of these two types of cluster data with standard priors on $h$ and $\Omegab h^2$ yields a competitive constraint on dark energy, $w=-1.06 \pm 0.15$.}
  \label{fig:clusterw}
\end{figure}

\begin{figure*}
  \centering
  \includegraphics[scale=\figscl]{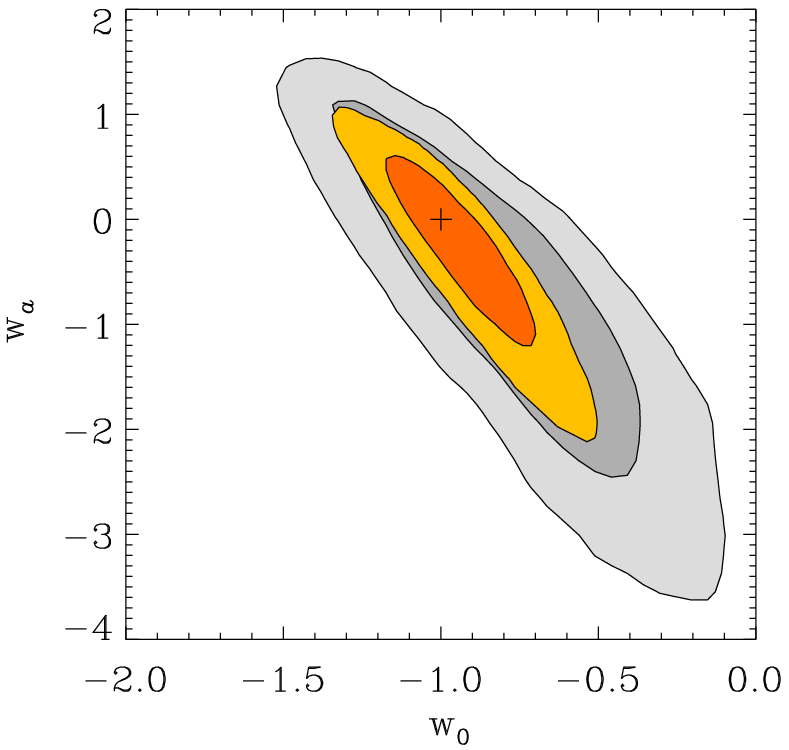}
  \hspace{1cm}
  \includegraphics[scale=\figscl]{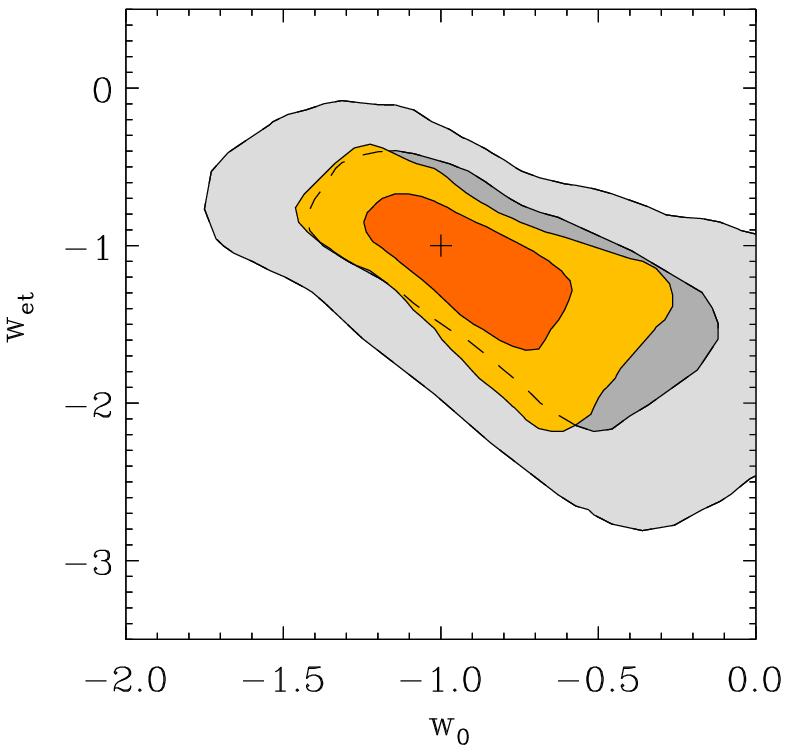}
  \caption{Joint 68.3 and 95.4 per cent confidence regions, including conservative systematic allowances, for parameters of the evolving $w$ models. Left: constraints on the $(w_0,w_a)$ model (\eqnref{eq:lindermod}) from the combination of XLF and 5-year WMAP data are shown in gray, as well as the combination of XLF, WMAP5 \citep{Dunkley09}, cluster \fgas{} \citepalias{Allen08}, SNIa \citep{Kowalski08} and BAO \citep{Percival07} data (gold). Right: Constraints on the $(w_0,\wet)$ model (\eqnref{eq:wevolmod}). The transition scale factor is marginalized over the range $0.5<\mysub{a}{t}<0.95$. Crosses in each panel indicate the \LCDM{} model, with constant $w=-1$.}
  \label{fig:evolvingw}
\end{figure*}

For models in which the dark energy equation of state, $w$, is constant in time, the joint constraints on \Omegam{} and $w$ from the XLF are shown in the left panel of \figref{fig:constantw} (purple contours). The constraint on the equation of state from the XLF alone, which is driven by the evolution in the cluster mass function, is $w=-1.0 \pm 0.2$ (including systematics), a factor of 3 improvement over \citetalias{Mantz08}. Our results on \Omegam{} and $\sigma_8$ are nearly identical to the \LCDM{} case (\tabref{tab:results}); these constraints are driven by the cluster data at low redshift and are thus largely insensitive to the properties of dark energy. \figref{fig:constantw} also demonstrates the agreement of our results with independent constraints from other cosmological data (\secref{sec:data}), including WMAP5 (blue), SNIa (green), cluster \fgas{} (red) and BAO (brown). The combination of these data sets appears as gold contours in the figure; the improved, marginalized constraints from this combination are $\Omegam=0.27 \pm 0.02$, $\sigma_8=0.79 \pm 0.03$ and $w=-0.96 \pm 0.06$. Our results are also consistent with those derived independently by \citet{Vikhlinin09a} from X-ray flux-selected clusters.

The constraints in the $\sigma_8$-$w$ plane from the XLF and WMAP5 data are shown in the right panel of \figref{fig:constantw}. The complementarity of the two data sets is evident, with the XLF constraint on $\sigma_8$ breaking an important degeneracy affecting the CMB data. The combination of only the XLF and WMAP5 data yields the grey contours in the figure, corresponding to the one-dimensional constraints $\Omegam=0.27 \pm 0.04$, $\sigma_8=0.78 \pm 0.04$ and $w=-0.95 \pm 0.14$.

We note that the combination of the two types of cluster data used here, XLF and \fgas{} (with standard priors on $h$ and $\Omegab h^2$), also places competitive constraints on dark energy: $w=-1.06 \pm 0.15$ (\figref{fig:clusterw}). Interestingly, this result is comparable to that of the combination of XLF with WMAP5 data, or \fgas{} with 3-year WMAP, Cosmic Background Imager,  Arcminute Cosmology Bolometer Array Receiver, and BOOMERanG data \citepalias[$w=-1.00 \pm 0.14$;][]{Allen08}.

As discussed in \secref{sec:cosmodel}, our evaluation of the linear matter power spectrum for models with $w \neq -1$ includes the effects of dark energy density perturbations. For comparison with other works where dark energy perturbations are neglected, we also performed an analysis where dark energy affects only the expansion of space (i.e. where it has uniform density) despite not being a cosmological constant. In this case, we obtain from the XLF the slightly larger constraint $w=-0.97 \pm 0.25$, including all systematic uncertainties in the usual way.

\subsection{Constraints on evolving $w$ models}

\begin{figure}
  \centering
  \includegraphics[scale=\figscl]{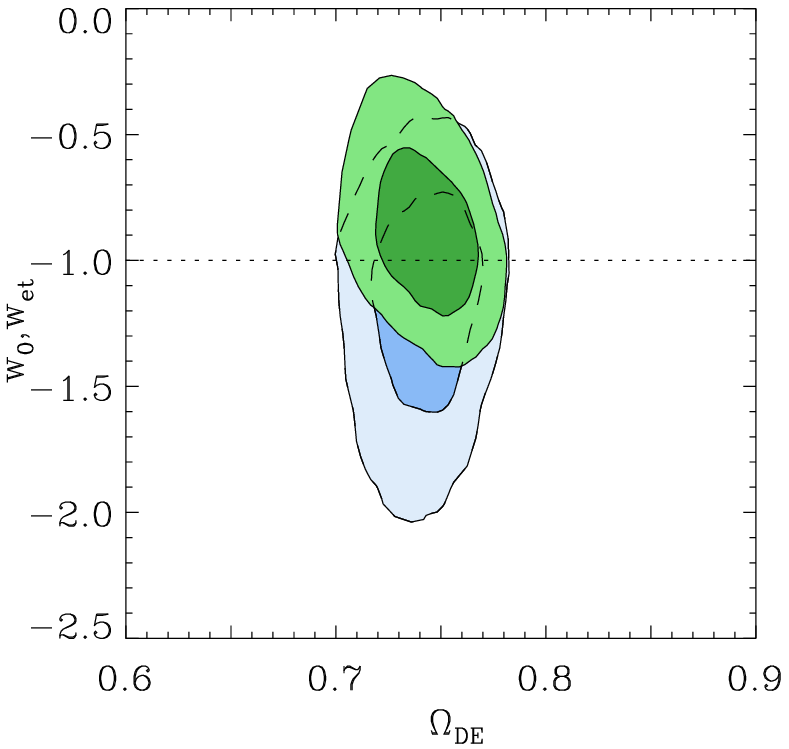}
  \caption{Joint 68.3 and 95.4 per cent confidence regions from the combination of XLF, \fgas{}, WMAP5, SNIa and BAO data, including systematic uncertainties, for parameters of the $(w_0,\wet)$ model (\eqnref{eq:wevolmod}). The results are the same as in the right panel of \figref{fig:evolvingw}, but with both $w_0$ (green contours) and \wet{} (blue) shown on the $y$-axis, against the dark energy density, $\mysub{\Omega}{DE}$, on the $x$-axis. The transition scale factor is marginalized over the range $0.5<\mysub{a}{t}<0.95$. The horizontal, dotted line indicates the cosmological constant model ($w_0=\wet=-1$).}
  \label{fig:evolww}
\end{figure}

Constraints on the two evolving $w$ models discussed in \secref{sec:cosmodel} (\eqnrefs{eq:lindermod} and \ref{eq:wevolmod}), $(w_0,w_a)$ and $(w_0,\wet)$, are shown respectively in the left and right panels of \figref{fig:evolvingw}. Results from the combination of XLF and WMAP5 data are shown as gray contours, and the combination of those two with \fgas{}, SNIa and BAO data is shown in gold. In both cases, there is no evidence for evolution in the dark energy equation of state, and the results are consistent with the cosmological constant model (indicated by a cross in the figure). Marginalized constraints are listed in \tabref{tab:results}. For the more general $(w_0,\wet)$ model, we find $w_0 = -0.88 \pm 0.21$ and $\wet = -1.05^{+0.20}_{-0.36}$ from the combination of all the data (\figref{fig:evolww}). These results are a significant improvement over previous constraints on this model from the combination of \fgas{}, CMB and SNIa data \citepalias[$w_0 = -1.05^{+0.31}_{-0.26}$, $\wet = -0.83^{+0.48}_{-0.43}$;][]{Allen08}.

Adding spatial curvature as a free parameter and fixing the transition redshift to 1 (\eqnref{eq:lindermod}), we obtain constraints equivalent to $\left[\sigma(\mysub{w}{p})\sigma(w_a)\right]^{-1}=15.5$, using the notation and definitions of the Dark Energy Task Force \citep[DETF;][]{Albrecht06}.

\section{Investigation of systematics} \label{sec:systematics}

\subsection{Sensitivity to priors} \label{sec:priors}

Our analysis of the XLF data alone includes priors to constrain the Hubble parameter and mean baryon density, and to marginalize over systematic allowances on the mass function, cluster sample completeness and cluster gas mass fraction (\tabref{tab:paramlist}). To investigate the influence of each prior individually, we importance sampled the Markov chains produced in the XLF analysis for the constant $w$ model, reducing the width of each prior in turn by a factor of two. If importance sampling a particular prior in this way changes the posterior distribution for parameters of interest, we can conclude that the prior is influential.

\begin{figure}
  \centering
  \includegraphics[scale=\figscl]{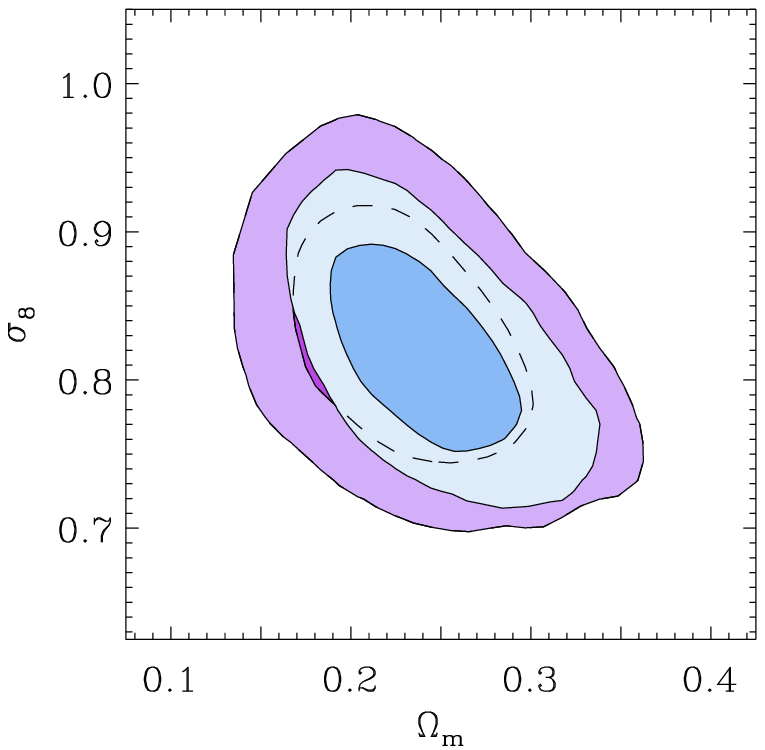}
  \caption{Joint 68.3 and 95.4 per cent confidence regions for \Omegam{} and $\sigma_8$ in the constant $w$ model from the XLF data, including the standard systematic allowances in \tabref{tab:paramlist}, are shown in purple. The blue contours result from reducing the width of all of the allowances associated with the \fgas{} model of \citetalias{Allen08} by a factor of two. The figure demonstrates that our results on \Omegam{} and $\sigma_8$ are limited by the systematic uncertainty in \fgas{}. (These allowances have no effect on the $w$ constraint.) Improvements in hydrodynamical simulations and the incorporation of gravitational lensing mass measurements offer the possibility of significantly reducing the uncertainty in \fgas{}, and thus improving the constraints.}
  \label{fig:fgasprior}
\end{figure}

Of the priors listed above, only the systematic allowances associated with the determination of \fgas{} are significant. Since the gas mass fraction determines the overall mass scale when \Mgas{} is used as a proxy for total mass, its systematic uncertainty affects primarily the constraints on \Omegam{} and $\sigma_8$, as shown in \figref{fig:fgasprior}. Following \citetalias{Allen08}, our standard analysis uses conservative systematic allowances for \Chandra{} calibration (10 per cent), non-thermal pressure support (10 per cent), the depletion of baryons in clusters with respect to the cosmic mean (20 per cent), and evolution with redshift in the baryon fraction and stellar content of clusters (10 and 20 per cent). The most significant for this work are on the depletion of baryons in clusters, which determines the width of the \Omegam{} constraint, and the amount of non-thermal pressure and overall calibration of \Chandra{} for temperature measurements, which determine the constraint on $\sigma_8$ at fixed \Omegam{}.

In contrast, none of the priors affects the constraint on $w$ significantly. This was confirmed by an additional analysis in which all of the systematic uncertainties were eliminated, i.e. every nuisance parameter was fixed rather than being marginalized over. Ordinarily, this would indicate that our constraints on dark energy are statistically limited, and this may plausibly be the case; however, we note that our constraints may also be limited by the fact that the effect of dark energy perturbations on the mass function is not yet understood. If these perturbations produce changes in the high-mass tail of the mass function at the few tens of per cent level, there may be additional constraining power available from the current XLF data.

\begin{table*}
  \centering
  \caption{Marginalized 68.3 per cent confidence intervals on cosmological parameters from our analysis, including all systematic uncertainties in \tabref{tab:paramlist}. The various cosmological models are discussed in \secref{sec:cosmodel}. ``All'' refers to the combination of XLF, cluster \fgas{}, WMAP5, SNIa and BAO data.}
  \begin{tabular}{cccccc}
    \hline
    Data & Model & \Omegam{} & $\sigma_8$ & $w,w_0$ & $w_a,\wet$ \\
    \hline
    XLF       & \LCDM{} & $0.23 \pm 0.04$ & $0.82 \pm 0.05$ & --- & --- \\
    XLF+WMAP5 & \LCDM{} & $0.26 \pm 0.02$ & $0.80 \pm 0.02$ & --- & --- \\
    all       & \LCDM{} & $0.257 \pm 0.015$ & $0.80 \pm 0.02$ & --- & ---\vspace{2mm} \\
    XLF       & constant $w$ & $0.23 \pm 0.04$ & $0.82 \pm 0.05$ & $-1.01 \pm 0.20$ & --- \\
    XLF+WMAP5 & constant $w$ & $0.27 \pm 0.04$ & $0.78 \pm 0.04$ & $-0.95 \pm 0.14$ & --- \\
    XLF+\fgas & constant $w$ & $0.22 \pm 0.04$ & $0.83 \pm 0.05$ & $-1.06 \pm 0.15$ & --- \\
    all       & constant $w$ & $0.272 \pm 0.016$ & $0.79 \pm 0.03$ & $-0.96 \pm 0.06$ & ---\vspace{2mm} \\
    XLF+WMAP5 & evolving $w$ ($w_a$)  & $0.26 \pm 0.04$ & $0.79 \pm 0.05$ & $-0.77 \pm 0.31$ & $-0.34^{+0.72}_{-1.42}$\vspace{1mm} \\
    all       & evolving $w$ ($w_a$)  & $0.256 \pm 0.016$ & $0.80 \pm 0.03$ & $-0.93 \pm 0.16$ & $-0.13^{+0.47}_{-0.73}$\vspace{2mm} \\
    XLF+WMAP5 & evolving $w$ ($\wet$) & $0.26 \pm 0.04$ & $0.79 \pm 0.05$ & $-0.73 \pm 0.40$ & $-1.10^{+0.59}_{-0.39}$\vspace{1mm} \\
    all       & evolving $w$ ($\wet$) & $0.257 \pm 0.016$ & $0.80 \pm 0.03$ & $-0.88 \pm 0.21$ & $-1.05^{+0.20}_{-0.36}$\\
    \hline
  \end{tabular}
  \label{tab:results}
\end{table*}

\subsection{Choice of mass function} \label{sec:mfcnpriors}

In our standard analysis, we have used the mass function of \citet{Tinker08} with cluster radius defined by an overdensity of $300\Omegam(z)$. One could also justify using the mass function determined at a cluster radius closer to the radius where the follow-up mass measurements are made, $r_{500}$. We therefore repeated the XLF analysis using the mass function given by Tinker et~al. for overdensity $1600\Omegam(z)$. For \Omegam{} values of 0.2--0.3 and $z<0.5$, this corresponds to overdensities of 300--1000, compared with 50--200 for $300\Omegam(z)$. Our results with this alternative mass function were virtually identical to the standard results.

Similarly, we repeated the analysis using the older mass function of \citet{Jenkins01} at overdensity $324\Omegam(z)$. Here we used a 20 per cent Gaussian systematic allowance  on only the normalization of the mass function and fixed the shape, as in \citetalias{Mantz08}. Again, the cosmological results were essentially identical to our standard results, despite the fact that the Jenkins function does not include evolution with redshift, while the Tinker function does. This indifference to the details of the mass function is consistent with the observation in \secref{sec:priors} that the current results are insensitive to the systematic allowances on the mass function and its evolution.

\subsection{Note on \Chandra{} calibration} \label{sec:chandracal}

A major, recent (21 January 2009) update to the \Chandra{} ACIS effective area at soft energies is accounted for in our analysis of follow-up \Chandra{} observations for the XLF data (see \scalingpaper{}). However, the raw data analysis in \citetalias{Allen08} predates this calibration update, which in our tests typically results in an increase in the inferred gas fraction of $\sim 10$ per cent. We have accounted for this correction by shifting the center of the ``\Chandra{} calibration'' nuisance parameter in the \citetalias{Allen08} model such that the preferred value of \fgas{} increases correspondingly by 10 per cent. The Gaussian prior on this parameter also has a width of 10 per cent, so this systematic allowance encompasses both the old value and the new value expected from the calibration update.

The effect of this higher gas mass fraction is to shift the \Omegam{} constraint from the \fgas{} analysis to lower values by approximately 0.02. A corresponding shift appears in the XLF results, since the \Omegam{} constraints are partially driven by our use of a subset of the \fgas{} data to set the overall cluster mass scale. The correction additionally results in a shift of similar magnitude towards lower $\sigma_8$ values, but has no effect on the determination of $w$. While we expect that the 10 per cent Gaussian allowance adequately reflects the systematic uncertainty in the overall \Chandra{} calibration, these trends should be kept in mind in interpreting our results. For example, if we have overestimated the effect that the calibration update has on the value of \fgas{}, our best fit results will shift slightly to higher \Omegam{} and $\sigma_8$.

\section{Conclusion} \label{sec:conclusion}

We have presented cosmological constraints obtained from an X-ray flux-limited sample of 238 massive galaxy clusters spanning the redshift range $z<0.5$. Follow-up \Chandra{} or \ROSAT{} X-ray observations of 94 of these clusters are incorporated, as detailed in \scalingpaper{}. Our analysis accounts for all selection biases, includes conservative allowances for systematic uncertainties, and, for the first time, produces simultaneous constraints on cosmology and the cluster scaling relations using a rigorous and fully self-consistent statistical method. The incorporation of follow-up data and our improved analysis method result in cosmological constraints that are a factor of 2--3 better than our previous results in \citetalias{Mantz08}, which were based on the same flux-limited sample of clusters. The results presented here are among the tightest and most robust constraints on cosmological parameters available from current data.

The constraints on spatially flat, cosmological constant models from our XLF data are $\Omegam=0.23 \pm 0.04$ and $\sigma_8=0.82 \pm 0.05$. Introducing a constant dark energy equation of state, $w$, as a free parameter, we find $w=-1.0 \pm 0.2$, obtaining the same tight constraints on \Omegam{} and $\sigma_8$. These results confirm at higher precision the first constraints on the dark energy equation of state from the analysis of XLF data reported by \citetalias{Mantz08}, and are consistent with independent findings based on cluster gas mass fractions, CMB anisotropies, type Ia supernovae, baryon acoustic oscillations, galaxy redshift surveys, and cosmic shear. We also find good agreement with recent, independent analyses of X-ray selected galaxy clusters \citep{Henry09,Vikhlinin09a} and optically selected clusters \citep{Rozo10}.

Using the combination of XLF, \fgas{}, WMAP5, SNIa and BAO data, we investigate evolving $w$ models, including the standard $(w_0,w_a)$ model as well as a $(w_0,\wet)$ model in which the transition scale factor, $\mysub{a}{t}$, is marginalized over. In both cases, the best current cosmological data remain consistent with a constant value of $w=-1$. The constraints on the non-flat $(w_0,w_a)$ model from the combination of current data are equivalent to DETF figure of merit $\left[\sigma(\mysub{w}{p})\sigma(w_a)\right]^{-1}=15.5$. Compared with the figure of merit of 8.3 reported by \citet{Wang08} from the combination of WMAP5, SNIa and BAO data, this result highlights the contribution of cluster data (XLF and \fgas{}), and their complementarity to other cosmological information.

Currently, the most significant systematic limitation on our results for \Omegam{} and $\sigma_8$ is systematic uncertainty in the cluster gas mass fraction, \fgas{}. Specifically, the dominant uncertainty in the determination of \Omegam{} is in the depletion of baryons in clusters relative to the cosmic mean, for which a 20 per cent systematic allowance is adopted in the present work. The prospect of significant improvement here exists, as numerical simulations incorporate more complete treatments of baryonic physics, especially feedback and star formation processes in the centers of clusters. The results on $\sigma_8$ are limited by our knowledge of the non-thermal pressure in clusters and the overall calibration of X-ray temperature measurements; cluster mass measurements via gravitational lensing provide an observational means of circumventing this issue, by directly constraining the overall mass scale of the data.

XLF results on dark energy remain statistically limited. The incorporation of additional X-ray, SZ (e.g. Planck, the South Pole Telescope and the Atacama Cosmology Telescope) and optical/near-infrared (e.g. the Dark Energy Survey and Pan-STARS) cluster surveys could lead to significant near-term improvements, although we stress that these surveys must have well-understood selection functions, sampling models, and scaling relations. The possibility also remains that cosmological simulations of the mass function including the effects of dark energy perturbations will reveal additional constraining power. It is therefore important to pursue a theoretical understanding of the effects of fluid dark energy on nonlinear structure formation, even as much larger X-ray (eROSITA\footnote{\url{http://www.mpe.mpg.de/projects.html\#erosita}}), and optical (e.g. the Large Synoptic Survey Telescope) cluster surveys promise to increase the data available for cluster cosmology by orders of magnitude.

We note that the statistical approach described in this work has significance for the planning of follow-up observations for future cluster surveys. Our method permits the use of partial follow-up, allowing scaling relations to be directly and robustly constrained over a wide range in mass without raising the cost in exposure time prohibitively, as would be the case for a complete, exhaustive follow-up campaign. At the same time, the incorporation of follow-up data provides a significant advantage over pure self-calibration approaches. In detail, any plan for X-ray follow-up observations of future surveys should also account for the additional benefit of observing dynamically relaxed clusters suitable for the \fgas{} test; as we have shown, the combination of cluster \fgas{} data, which directly probe the expansion of the Universe, and the XLF, which measures the growth of structure, can provide precise cosmological constraints independent of CMB and SNIa observations.

Readers interested in the simultaneous constraints on cluster scaling relations produced in this work should refer to \scalingpaper{}. MCMC samples encoding the results of these two papers will be made available for download on the web.\footnote{\url{http://www.stanford.edu/group/xoc/papers/xlf2009.html}}

\section*{Acknowledgments} \label{sec:acknowledgements}

We thank the reviewer for a very careful reading of the manuscript, and Jeremy Tinker for sharing details of his mass function work. We are also grateful to Glenn Morris, Stuart Marshall and the SLAC unix support team for technical support. Calculations were carried out using the KIPAC XOC and Orange compute clusters at the SLAC National Accelerator Laboratory and the SLAC Unix compute farm. We acknowledge support from the National Aeronautics and Space Administration (NASA) through LTSA grant NAG5-8253, and though Chandra Award Numbers DD5-6031X, GO2-3168X, GO2-3157X, GO3-4164X, GO3-4157X, GO5-6133, GO7-8125X and GO8-9118X, issued by the Chandra X-ray Observatory Center, which is operated by the Smithsonian Astrophysical Observatory for and on behalf of NASA under contract NAS8-03060. This work was supported in part by the U.S. Department of Energy under contract number DE-AC02-76SF00515. AM was supported by a William~R. and Sara Hart Kimball Stanford Graduate Fellowship.

\bibliographystyle{mnras}
\def \aap {A\&A} 
\def \statisci {Statis. Sci.}
\def \physrep {Phys. Rep.}
\def \pre {Phys.\ Rev.\ E}
\def \sjos {Scand. J. Statis.} 
\def \jrssb {J. Roy. Statist. Soc. B} 
\def \pan {Phys. Atom. Nucl.} 
\def \epja {Eur. Phys. J. A} 
\def \epjc {Eur. Phys. J. C} 
\def \jcap {J. Cosmology Astropart. Phys.} 
\def \ijmpd {Int.\ J.\ Mod.\ Phys.\ D}
\def \araa {ARA\&A}
\def \aj {AJ}
\def \apj {ApJ}
\def \apjl {ApJL}
\def \apjs {ApJS}
\def \mnras {MNRAS}
\def \nat {Nat}
\def \pasj {PASJ}
\def \gca {Geochim.\ Cosmochim.\ Acta}
\def \npa {Nucl.\ Phys.\ A}
\def \plb {Phys.\ Lett.\ B}
\def \prc {Phys.\ Rev.\ C}
\def \prd {Phys.\ Rev.\ D}
\def \prl {Phys.\ Rev.\ Lett.}

\bsp
\label{lastpage}
\end{document}